\documentclass{optica-article}

\journal{opticajournal} 

\articletype{Research Article}

\usepackage{lineno}
\usepackage{todonotes}
\usepackage[section]{placeins}
\usepackage{siunitx}
\DeclareSIUnit\angstrom{\text {Å}}
\DeclareSIUnit\electron{\mathrm{e}^{-}}

\begin{document}

\title{Fast Ptychographic Near-Field Computed Tomography}

\author{Sami Wirtensohn\authormark{1,2,3*},  Silja Flenner\authormark{2}, Imke Greving\authormark{2}, Jens Brehling\authormark{2}, Hilmar Burmester\authormark{2}, Dominik John\authormark{1,2}, Sara Baggio\authormark{1}, Franziska Hinterdobler\authormark{1}, Maximiliane Wojke\authormark{1}, Kritika Singh\authormark{2}, Benedikt J. Daurer\authormark{4}, Johannes Hagemann\authormark{3}, Frank Seiboth\authormark{3}, Sara Savatovi\'c\authormark{1}, Fritz Vollrath\authormark{5}, Julia Herzen\authormark{1}}

\address{
\authormark{1}Research Group Biomedical Imaging Physics, Department of Physics, TUM School of Natural Sciences \& Munich Institute of Biomedical Engineering, Technical University of Munich, James-Franck-Straße 1, 85748 Garching, Germany\\
\authormark{2}Institute of Materials Physics, Helmholtz-Zentrum Hereon, Max-Planck-Straße 1, 21502 Geesthacht, Germany\\
\authormark{3}Centre for X-Ray and Nano Science CXNS, Deutsches Elektronen-Synchrotron (DESY), Notkestraße 85, 22607 Hamburg, Germany\\
\authormark{4}Diamond Light Source Ltd, Fermi Ave, Didcot OX11 0DE, United Kingdom\\
\authormark{5}Department of Biology, University of Oxford, South Parks Road, Oxford OX1 3EL, United Kingdom
}

\email{\authormark{*}sami.wirtensohn@tum.de}

\begin{abstract*}
Near-field X-ray ptychography enables quantitative three-dimensional imaging with nanometer-scale resolution, but its broader application is limited by long acquisition times -- often necessitating cryogenic or vacuum environments -- and demanding sample preparation procedures.
In this paper, we present a flexible near-field ptychography setup that operates under ambient conditions and features an adjustable geometry for different sample sizes and resolution requirements.
By replacing stepwise angular acquisition with a fly-rotation sample motion, the scanning overhead is reduced by a factor of 11, drastically lowering the tomographic scan time. To address the bottleneck in sample preparation, we also introduce a sample milling machine that enables fast preparation of specimen pillars with diameters down to \SI{20}{\micro\meter}.
By lowering both preparation effort and measurement time while maintaining high spatial resolution, the presented system overcomes key limitations of near-field ptychography and makes quantitative X-ray nanotomography substantially more accessible for biological research.
\end{abstract*}


\section{Introduction}
To fully understand the structure and functioning of biological samples, it is crucial to observe the system in its native three-dimensional context~\cite{Varga2015, Walton2015, Nathansen2024, RajaSomu2025}. Due to its high resolution and non-destructive nature, X-ray full-field nanotomography became a valuable research tool. Since soft tissues typically provide low-attenuation contrast, making the identification of internal structures difficult \cite{David2002, Berthe2024}, several approaches have been developed to retrieve the phase shift induced by the sample. The phase signal provides enhanced soft tissue contrast and directly reflects the specimen's electron density~\cite{Fitzgerald2000, Schaff2020}.

A widely used approach to exploit the contrast enhancement of phase effects in full-field imaging is Zernike phase-contrast transmission X-ray microscopy (ZPC-TXM)~\cite{Zernike1942, Schmahl1994}. It consists of two optics, one upstream and one downstream of the sample, creating a strong magnification. To visualize phase effects, a phase-shifting ring is added in the back focal plane of the second optic. It creates a $\frac{\pi}{2}$-phase offset between the non-sample-refracted and the sample-refracted part of the beam~\cite{Neuhausler2003}. This offset leads to interference effects in the detector plane, creating a contrast boost~\cite{Andrews2008, Stampanoni2010, Storm2020}. While this is a robust approach, it comes with two downsides. First, it does not give access to the quantitative phase information. Hence, the exact phase shift due to the sample remains unknown. Second, the objective lens downstream of the sample drastically reduces the setup's efficiency~\cite{Vartiainen2014}, thereby leading to a high sample dose. As a result, ZPC-TXM is less suitable for dose-sensitive samples such as soft tissue, where degeneration may occur, ultimately leading to structural changes and sample displacement.

More dose-sensitive approaches are propagation-based methods. These methods retrieve the phase distribution by modeling wave propagation using the measured intensity pattern on the detector and the known propagation distance. The magnification is achieved using a cone-beam geometry, which naturally magnifies the specimen. A powerful and straightforward approach of this kind is the Paganin method~\cite{Paganin2002}. It assumes a homogeneous sample and couples attenuation and phase, thereby allowing the extraction of the phase map on a single projection~\cite{Paganin2002}. Due to its simplicity, it has found wide application~\cite{Walsh2021, Twengstrm2022, Schaeper2025}. However, since the modalities are coupled, it cannot retrieve quantitative attenuation and phase information simultaneously and depends on speckle-free, flat illumination.

To measure non-homogeneous specimens, scans at multiple distances can be acquired~\cite{Cloetens1999}. The additional information serves as constraints during signal retrieval, thereby relaxing the required approximations and enabling access to quantitative phase information~\cite{Cloetens1999}. This makes multi-distance holotomography an appealing imaging method where high accuracy matters~\cite{Yu2017, Monaco2022, Kalbfleisch2022, Nikitin2024}. However, the additional measurements prolong the scan time, thereby increasing the dose and the setup's stability requirements. It also necessitates more sophisticated registration and retrieval algorithms compared to single-distance approaches~\cite{Lucht2025} and reaches its limits in high-resolution quantitative phase imaging of complex, extended, or multiscale objects.

To overcome these limitations, near-field ptychography offers a promising alternative. In this method, the sample is scanned laterally across a structured illuminating beam while recording interference patterns at multiple positions~\cite{Stockmar2013}. The strong overlap constraint between adjacent patterns, together with the precisely known scan positions and geometry, enables iterative retrieval of the complex wave functions of both the probe (illumination) and the object (sample)~\cite{Thibault2008, Enders2016}. Hence, near-field ptychography provides quantitative access to both phase and attenuation signals at nanometer-scale resolution, enabling the decomposition of a scanned specimen into two base materials~\cite{Taphorn2022, John2026Adv}.

Despite its advantages, near-field ptychography is constrained by long tomographic acquisition times~\cite{Dierolf2010, Taphorn2022, Shirani2024}. At each projection angle, the specimen must be scanned over many lateral positions. This results in significant waiting time for motor movement, which is further referred to as \textit{overhead}. These extended scan times not only reduce experimental throughput but also further increase the dose and impose stringent stability requirements on the optical setup and sample environment. For biological specimens, which are highly susceptible to motion and radiation damage, this poses a particular challenge. 
While cryogenic environments have been employed to mitigate radiation and motion effects -- such as the cryogenic vacuum sample chamber at the cSAXS beamline of the Swiss Light Source, which enables results of high quality and resolution~\cite{Shahmoradian2017, Holler2018} -- the speed bottleneck remains.

Consequently, current methods pose a compromise between fast, low-dose imaging and robust, quantitative phase retrieval. Reducing the total scan time of near-field ptychography is therefore crucial to break the speed-accuracy trade-off and enable quantitative, high-throughput phase-contrast imaging of complex, dose-sensitive specimens across scientific fields such as materials science, medicine, and biology.

Near-field ptychography further imposes two intrinsic constraints on the sample size. First, self-interference within the sample sets a thickness limit: if the specimen exceeds a certain threshold, the thin sample approximation becomes invalid and multislice modeling is required to reconstruct the wavefield correctly~\cite{Stockmar2013, Tsai2016, Hu2023}. Second, when the outermost interference fringes of the sample extend beyond the detector, high spatial frequency information is lost, leading to a degradation of resolution in the peripheral areas.
These requirements translate into strict size and shape constraints during sample preparation for nanoimaging and are  key to achieving high-quality, high-resolution scans. Therefore, sample preparation remains challenging and requires particular care.

Plasma-focused ion beam (FIB) milling has therefore become an important technique for preparing samples~\cite{Lombardo2012}, as it allows precise cutting of conducting samples for nanotomography. It furthermore enables the free selection of the extraction point and the desired cutting geometry. However, the process is generally time-consuming, especially for non-conducting samples with high carbon content, such as biological specimens. Additionally, FIB is typically limited to sizes below \SI{100}{\micro\meter} due to redeposition bottlenecks~\cite{Mayr2021} and the critical lift-out process~\cite{Lombardo2012}.
During the lift-out process, a metal manipulator is attached to the top of the sample pillar. The sample pillar is subsequently cut from the main sample body by the FIB at its base and can then be transferred to the desired mounting system for the experiment. However, an improper base cut will either lead to a break or bending of the manipulator or even loss of the sample. Due to the steep viewing angle of the electron beam in FIB-SEM systems, the interface between the sample pillar and the sample substrate is only partially visible during milling~\cite{Lombardo2012}. As a result, it is often unclear whether the base has been fully detached. To ensure full detachment, the milling process is typically extended beyond the estimated boundary of the base. A higher sample thickness extends the required milling time and makes it more difficult to visually judge whether the base has been completely separated. Therefore, a thicker sample reduces the chance of full detachment, making it more prone to failure. Furthermore, the cutting of biological materials can only be performed with limited current, since the heat generated during cutting can dry out the sample and alter its physical properties, such as increased brittleness~\cite{Wolff2018, Singh2023}. This can further extend the preparation time for biological samples up to \SI{48}{\hour}.
Together with high costs and the excessive milling times required for larger volumes, FIB entails substantial overall experimental effort, which poses a limitation to nanoimaging. This is particularly true for biological studies, where additionally a large number of samples is often required.

To address these challenges, we implement an improved sample milling machine, based on the concept of Holler et al. \cite{Holler2020}, that mechanically cuts biological specimens to diameters below \SI{50}{\micro\meter}, enabling fast, low-effort preparation across a broad range of pillar sizes. Furthermore, we present a near-field ptychography setup with a highly flexible geometry that enables a wide range of magnifications and sample sizes under ambient conditions. To drastically reduce scan times and facilitate high-throughput imaging, we introduce the concept of fly rotation to near-field ptychography, which allows the acquisition of interference patterns during continuous sample rotation. 

\section{Method}

\subsection{Samples}
This study utilizes a diverse set of reference samples, selected to evaluate performance across varying material classes and preparation methodologies. For friable materials, a critical point-dried sausage specimen was provided by the \textit{German Federal Institute for Risk Assessment}. To assess conventional embedding protocols, \textit{Helmholtz-Zentrum Hereon} supplied two additional specimens: a zebrafish section embedded in paraffin and a moss sample embedded in epoxy. The reference set further includes solid matrices with distinct structural characteristics. A commercially available wooden toothpick serves as the standard for plant-based cellulose. Separately, an ivory specimen was obtained from the tusk of a male African savannah elephant (Loxodonta africana) through the \textit{University of Oxford}. This sample was harvested approximately \SI{1}{\centi\meter} beneath the outer layer, with the Schreger lines clearly visible.

\subsection{Sample Milling Machine}
\begin{figure}[htbp]
    \centering
    \includegraphics[width=\linewidth]{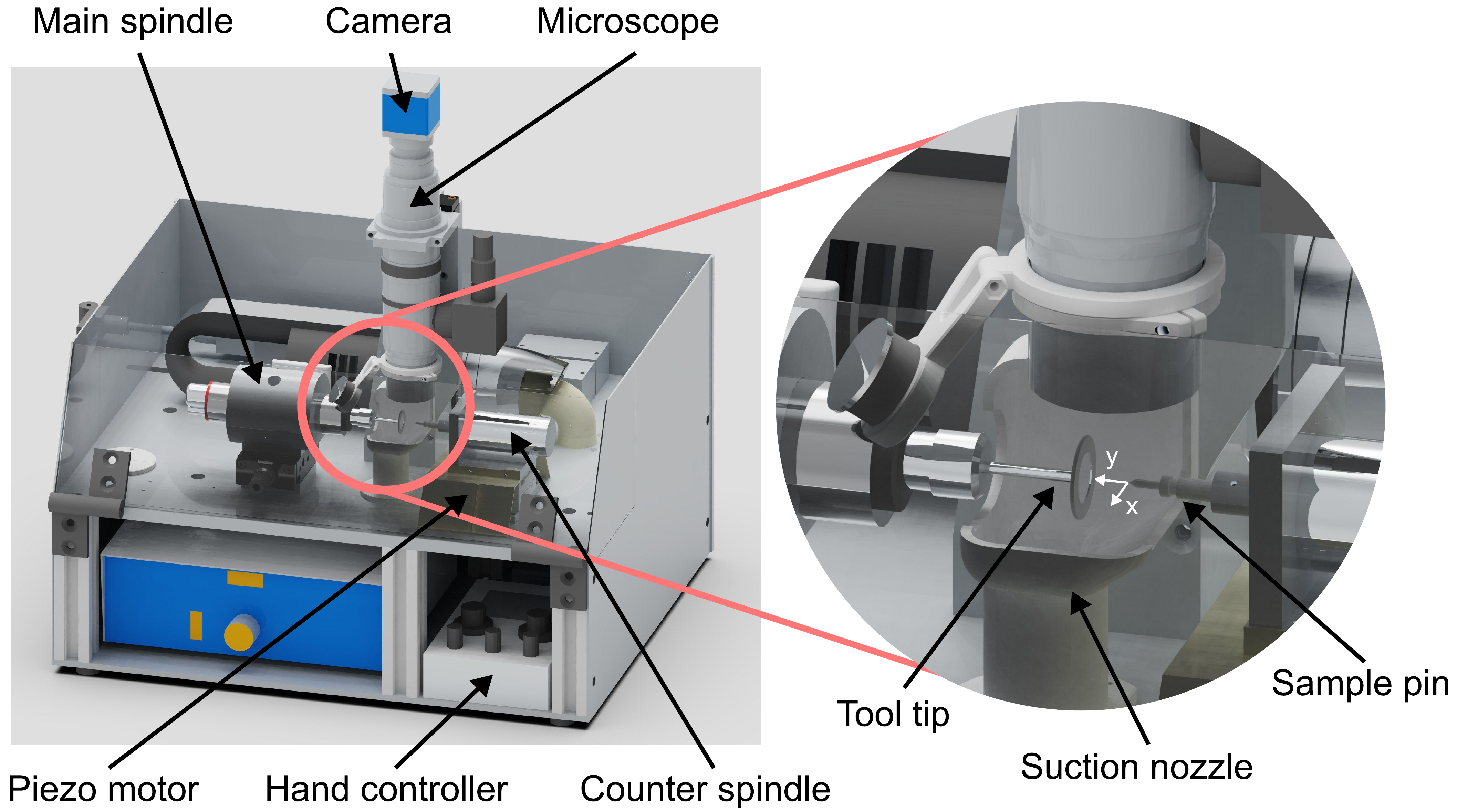}
    \caption{Rendering of the sample milling machine. The sample on the counter spindle rotates in the opposite direction to the cutting tool on the main spindle. The counter spindle with the sample on top can be moved precisely by two piezoelectric actuators. The cutting process can be observed through a camera connected to an optical light microscope. Excessive dust and cutting waste are removed by a suction system.}
    \label{fig:lathe}
\end{figure}

The sample preparation for nanotomography comes with high requirements in size and shape of the specimens to meet the tight beam geometry of advanced synchrotron instruments. Depending on the imaging method, required resolution, and sample material, the specimens must be reduced to sizes between \SI{10}{\micro\meter} and \SI{200}{\micro\meter}. For that, the \textit{Helmholtz-Zentrum Hereon} operates a FIB, which enables precise cutting of conducting samples for nanotomography.
However, for non-conducting samples with high carbon content, the process is very time-consuming and is usually limited by the sample diameter, which poses challenges during the critical lift-out process.
To reduce the required time and to extend the capabilities of sample preparation, a sample milling machine was built to cut down samples mechanically to sizes required for nanotomography (current minimum diameter: \SI{20}{\micro\meter} for steel and \SI{35}{\micro\meter} for wood), making the beamline more accessible for users. It also has the capability to create samples with a diameter greater than \SI{100}{\micro\meter}, which is a limitation of FIB preparation.
Additionally, it can be used to quickly cut down large samples for precise FIB polishing as a secondary step.

The sample milling machine is based on two spindles as shown in Figure~\ref{fig:lathe}. The first spindle, the main spindle, rotates an exchangeable cutting tool at high rates and removes the material from the specimen. For this purpose, a Chopper~230~H (\textit{Nakanishi Jaeger GmbH}, Ober-Mörlen, Germany) is chosen. It features a high nominal speed of $50000$ rpm and a concentricity of the inner cone of less than \SI{1}{\micro\meter}. The second spindle, the counter spindle, rotates at a lower rate in the opposite direction and holds the workpiece. The counter spindle reduces cutting forces and leads to less tool deflection, resulting in improved precision. It also enhances the surface finish and can increase the relative cutting speed if required. Here, a DCX32 (B78B8F36350C) motor (\textit{maxon international ltd.}, Sachseln, Switzerland) is deployed. The counter spindle is mounted on two piezoelectric linear stages (\textit{SmarAct Holding GmbH}, Oldenburg, Germany), which can move the sample precisely in the horizontal xy-plane of the cutting tool. This allows for fine adjustments during the cutting process. At the current state, the lathe is controlled with a hand controller, and the cutting operation can be observed through a camera-based light microscope. It enables active monitoring of the sample size after pre-calibrating the pixel size. The light microscope is based on a VH-Z100 objective (\textit{KEYENCE DEUTSCHLAND GmbH}, Frankfurt am Main, Germany) and a MikroCam PRO HDMI Autofocus camera (\textit{Bresser GmbH}, Rhede, Germany). The sample milling machine is inspired by a lathe system of M. Holler et al. \cite{Holler2020, Messler2021}, which is already in operation at the cSAXS beamline at the Swiss Light Source.
The samples prepared by the sample milling machine shown in this work are cut with a \textit{Garant} Z5 -- fine AS0820 burr (\textit{Hoffmann Groupe}, Munich, Germany), made out of carbide. However, the tool can be replaced by any commercially available cutting or grinding tool.

\subsection{Near-Field Ptychography}
\begin{figure}[htbp]
    \centering
    \includegraphics[width=\linewidth]{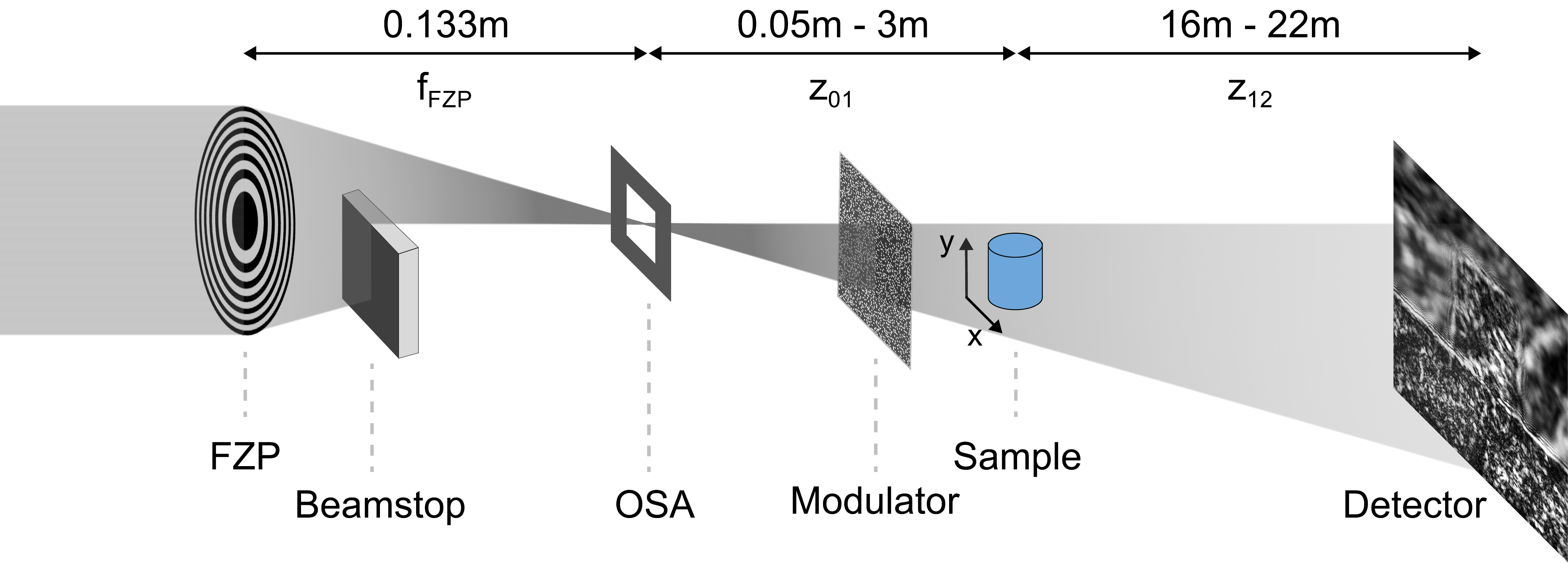}
    \caption{Schematic of the near-field ptychography setup, with the focus-sample distance $z_{01}$ and the sample detector distance $z_{12}$. A Fresnel Zone Plate (FZP) with the focal distance $f_{\text{FZP}}$ creates a cone beam, which is used to geometrically magnify the sample by a long propagation distance. To increase the variation in the probe, a speckle pattern is placed into the beam between the FZP and the sample. The sample is stepped in x- and y-direction over multiple positions, while the sample-detector distance remains constant. At each sample position, an interference pattern is recorded.}
    \label{fig:pty_setup}
\end{figure}

The experiments are carried out at the nanotomography endstation of the P05 beamline, operated by the \textit{Helmholtz-Zentrum Hereon}, located in Hamburg, Germany, at \textit{DESY}, PETRA III. The X-ray beam is generated by a 2-meter-long U29 undulator and produces a source size of $\SI{34.6}{\micro\meter} \times \SI{6.3}{\micro\meter}$ with a divergence of $\SI{28.9}{\micro\radian} \times \SI{1.6}{\micro\radian}$~\cite{Flenner2022p, Wirtensohn2025}. The X-ray beam was adjusted to \SI{11}{\kilo\electronvolt} using a Si-channel-cut monochromator. For imaging, a \textit{Hamamatsu} C12849-101U detector (\textit{Hamamatsu Photonics K.K.}, Hamamatsu, Japan)  was employed, positioned approximately \SI{19}{\meter} downstream from the sample in an adjacent hutch. This detector features a 2048 × 2048 pixel array with \SI{6.5}{\micro\meter} pixel pitch and a 1:1 fiber-coupled \SI{10}{\micro\meter} thick Gadox scintillator, capturing 16-bit images~\cite{Flenner2020, Wirtensohn2024}.

For the near-field ptychography setup, a cone beam is created by a Fresnel Zone Plate (FZP) made out of gold, with a diameter of \SI{300}{\micro\meter} and an outermost zone width of \SI{50}{\nano\meter}, manufactured by the \textit{Paul Scherrer Institute}, Villigen, Switzerland. As shown in Figure~\ref{fig:pty_setup}, a beamstop covers slightly more than half of the Zone plate to block out the transmitted beam in the center of the FZP. The FZP has a focus distance of \SI{0.133}{\meter} and in its focal plane, an Order Sorting Aperture (OSA) is placed to block higher diffraction orders.

The sample is placed on a \textit{SmarAct} SmarPod 110.45 (\textit{SmarAct Holding GmbH}, Oldenburg, Germany), which is mounted on an air-bearing rotation stage RTU150 (\textit{LAB Motion Systems}, Bekkevoort, Belgium). Due to the air-bearing, the rotation stage shows high precision and stability, while the SmarPod allows the precise stepping of the sample, required for ptychography. At the P05 nanotomography endstation, the rotation stage is positioned on a granite slider, which can be moved several meters away from the focal spot. The sample is measured against air at room temperature, making fast sample changing possible while maintaining a low sample handling effort. This makes the setup very flexible with a wide range of magnifications and an adjustable field of view.
If desired, an environmental control cell can set the humidity and temperature to given values to keep the sample conditions consistent throughout the scan. Furthermore, a nanoindentation system, a micromanipulator, and a flow cell for corrosion tests are available for \textit{in-situ} operation~\cite{Reimers2023, RajaSomu2025, Nopens2025, Ulrich2025}. To improve efficiency and avoid air scattering, vacuum tubes are placed between the sliders.

\begin{figure}[htbp]
    \centering
    \includegraphics[width=\linewidth]{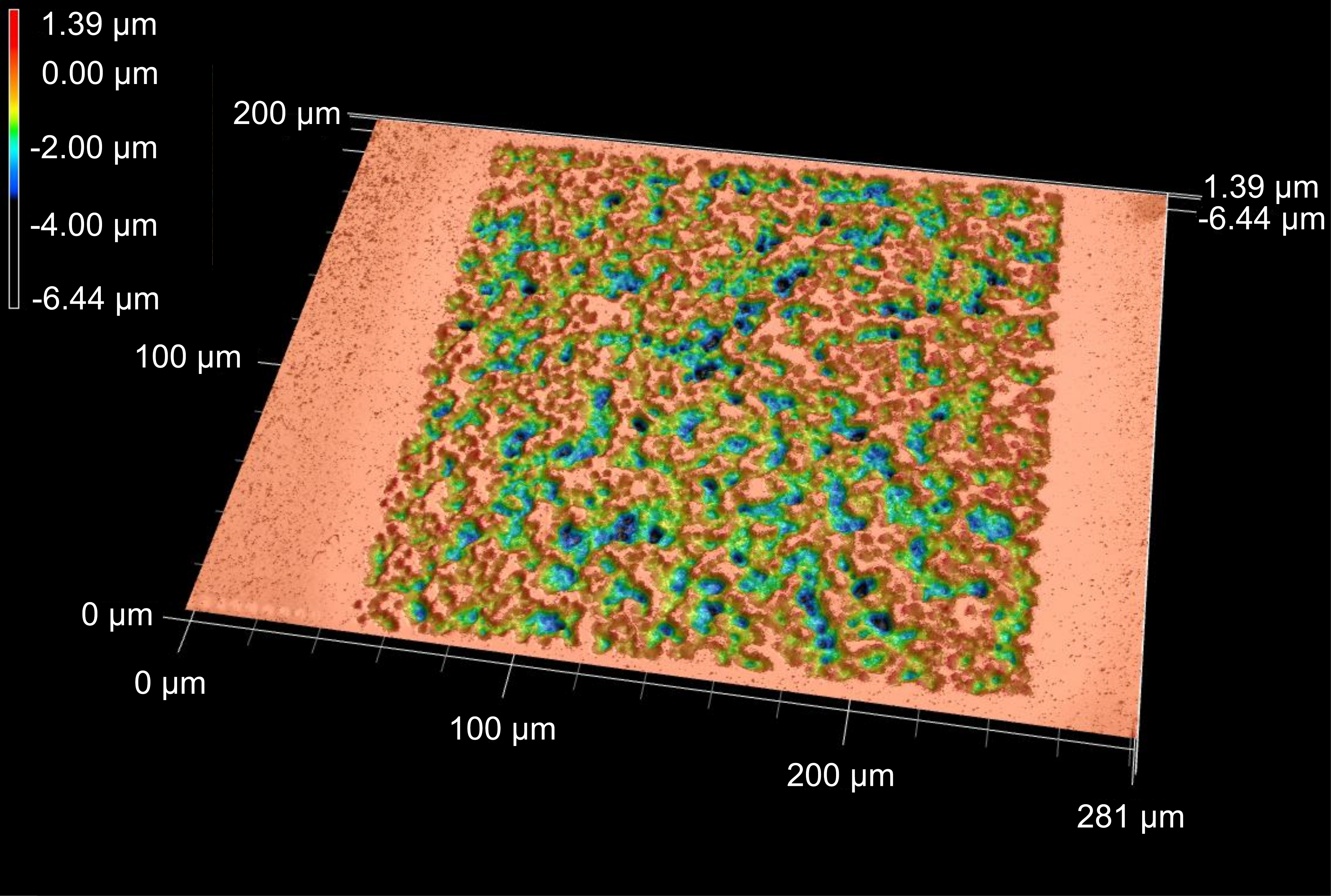}
    \caption{Surface plot of the diamond modulator plate. The surface structure is created by laser ablation and consists of randomly distributed round indentations with an approximate depth of \SI{4}{\micro\meter}. Random overlaps create locally stronger variations. The plate is written in an area of $\SI{200}{\micro\meter} \times \SI{200}{\micro\meter}$.}
    \label{fig:diamond_modulator}
\end{figure}

In near-field ptychography, a modulator is placed between the FZP and the sample to introduce strong variations in the probe, thereby improving convergence during signal retrieval. As standard sandpaper proved insufficiently stable, alternative
materials were investigated. Diamond plates were found to provide the needed long-term stability. Hence, a \SI{500}{\micro\meter} thick diamond plate is used as a modulator. To create wavefront variations in the diamond plate, round indentations are created by laser ablation with an approximate depth of \SI{4}{\micro\meter}, corresponding roughly to a phase shift of $\pi/2$ at an energy of \SI{11}{\kilo\electronvolt}. The holes are distributed randomly over the modulator's surface, allowing for overlap and creating locally stronger variations. The diamond plate is written in an area of $\SI{200}{\micro\meter} \times \SI{200}{\micro\meter}$. Its surface is shown in Figure~\ref{fig:diamond_modulator}.

\subsection{Scan Time Reduction}
To understand biological systems, it is not only important to retrieve high-quality data, but also to observe processes at different times. Furthermore, valid statements require a certain statistic, which translates into a high number of scans to be performed within limited beam time. Therefore, the scan time can not be neglected. To reduce the scan time of near-field ptychographic tomograms, we introduce the fly-rotation scan in the following.

\begin{figure}[htbp]
    \centering{}
    \includegraphics[width=\linewidth]{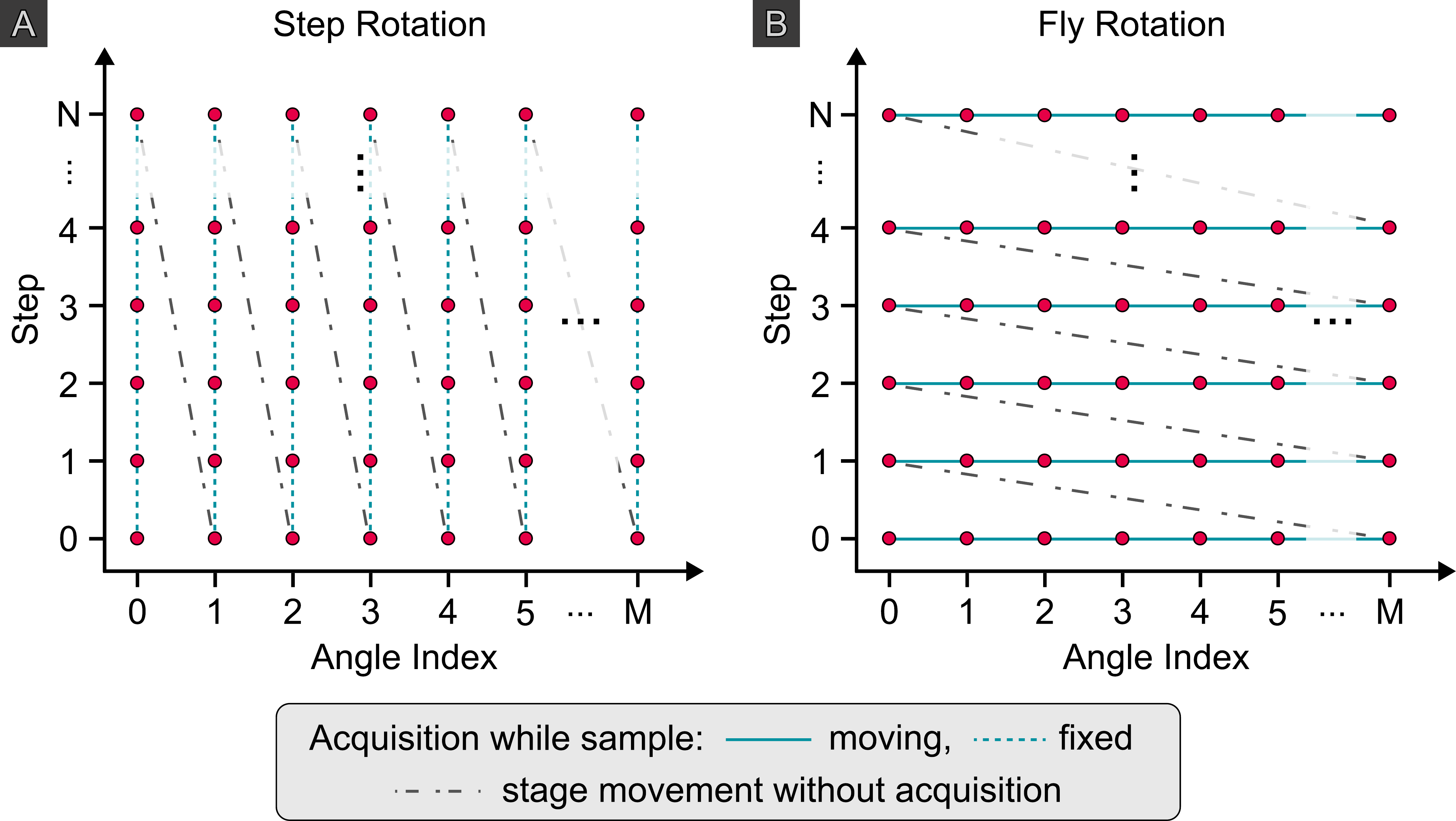}
    \caption{Schematic comparison between the step-rotation and fly-rotation scans. In the step-rotation scan (\textbf{A}), the sample is fixed during acquisition, and both motors (stepping motor and rotation axis) are moving in discrete steps. The overhead scales with $N \times M$, with $N$ the total number of steps and $M$ the total number of projection angles. During the fly-rotation scan (\textbf{B}), the rotation axis is moved continuously, while the stepping motor moves in discrete steps between the tomograms. Here, the overhead scales only with $N$.}
    \label{fig:scan_protocol}
\end{figure}
\subsubsection{Step Rotation}
A standard measurement protocol for near-field ptychograpic tomography is the step-rotation approach. At a fixed angle of the rotation stage, the sample is moved over $N$ stepping positions as visualized in Figure \ref{fig:scan_protocol} A. At each position, an interference pattern is recorded. After that, the next angle is approached and the process repeated until $M$ angles are measured, as required for a full tomogram. This approach has the advantage of high stability since the sample is static during image acquisition, and possible changes in the illumination are minimal because the time needed for a full set of interference patterns is kept short ($N \times$ exposure time). However, this approach of image acquisition is very time-consuming due to the long motor overhead. In addition, for high stability, a short waiting time is required after each motor movement to avoid sample vibrations from the acceleration. This significantly increases the required scan time for a full tomogram, since the overhead scales with $N \times M$.

\subsubsection{Fly Rotation}
To reduce the scan time, we introduce the concept of the fly rotation into near-field ptychography. In a fly-rotation scan, the sample is continuously rotated during image acquisition, drastically reducing the overhead. While this concept is already widely used for other tomography methods \cite{Wang2019, John2024, Lioliou2024}, it has not yet been demonstrated to work for near-field ptychography, due to the high level of positional precision required.
In a fly-rotation scan at each stepping position, a tomogram with continuous rotation is acquired, collecting all angles for one interference pattern position as shown in Figure~\ref{fig:scan_protocol}~B. After that, a new interference pattern position is reached by stepping the sample, and a second continuous rotating tomogram is started. This is repeated until all interference patterns are collected. By doing so, the motor overhead is considerably reduced, since the rotation stage is continuously moved while the sample stage is moved only once per interference pattern position. In a fly-rotation scan, the overhead therefore only scales with the number of steps $N$.

\subsection{Signal Retrieval \& Experimental Parameters}
Before the signal retrieval, all interference patterns are corrected for beam current, dark current, and bad pixels. Consequently, the corrected interference patterns are $2 \times 2$ binned and fed into \textit{PtyPy} for signal retrieval \cite{Enders2016}. The corrected mean flat-field is used as an initial guess for the probe. In all presented scans, the phase projections are retrieved by the difference map (DM) algorithm~\cite{Thibault2008, Giewekemeyer2010}. After signal retrieval, the phase projections are offset, ramp corrected, and unwrapped with \textit{toupy} \cite{Guizar-Sicairos2011, daSilvia2017}. The projections are then reconstructed with \textit{TomoPy} and the \textit{ASTRA Toolbox} by a Gridrec algorithm~\cite{Gursoy2014, Pelt2016} without a dedicated ring removal technique. The phase information is finally converted into electron densities~\cite{Birnbacher2021}.

The first near-field experiment consists of two tomographic scans. One was performed with step rotation and the other with fly rotation. For both scans, each projection consists of $16$ interference patterns, each exposed for \SI{0.5}{\second} arranged in a raster pattern~\cite{Clare2015, Baggio2026}. The scan parameters are chosen to keep the total number of projections close between the two scans. The exact numbers are listed in Table \ref{tab:scan_param}. Both scans are performed at a focus-sample distance of \SI{0.2538}{\meter} and a sample-detector distance of \SI{19.4413}{\meter}, resulting in an unbinned effective pixel size of \SI{83.76}{\nano\meter}. At the start of each scan, $50$ flat-field images are taken without the sample in the beam. During the scan protocol comparison, PETRA III ran in standard multi-bunch mode with a beam current of \SI{120}{\milli\ampere}.

The projections of both scans are retrieved by \textit{PtyPy} with $10000$ DM-iterations. After phase retrieval and correction, the projections are roughly aligned via a center-of-mass shift, and then horizontally and vertically aligned using tomo consistency implemented within \textit{toupy}.

In a second near-field experiment, the setup's speed capabilities and robustness are demonstrated by performing a second fly-rotation scan of the same object with a reduced number of projections and a lower exposure time of \SI{0.25}{\second} per interference pattern. Each projection also consists of $16$ interference patterns acquired in a raster pattern as before~\cite{Clare2015}. The scan is conducted at a different focus-sample distance of \SI{0.3000}{\meter} and a sample-detector distance of \SI{19.2633}{\meter}, resulting in an unbinned effective pixel size of \SI{99.68}{\nano\meter}. PETRA III ran in 40-bunch mode with a beam current of \SI{110}{\milli\ampere}, which is less optimal for near-field imaging due to higher beam fluctuations.

The data from the second experiment is processed as described above. However, the phase is retrieved with $5000$ iterations of the DM algorithm, and the alignment step is skipped. Hence, the data is directly fed into \textit{TomoPy} for volume reconstruction after applying the phase corrections.

\section{Experimental Results}

\subsection{Sample Milling Machine}
\begin{figure}[htbp]
    \centering{}
    \includegraphics[width=\linewidth]{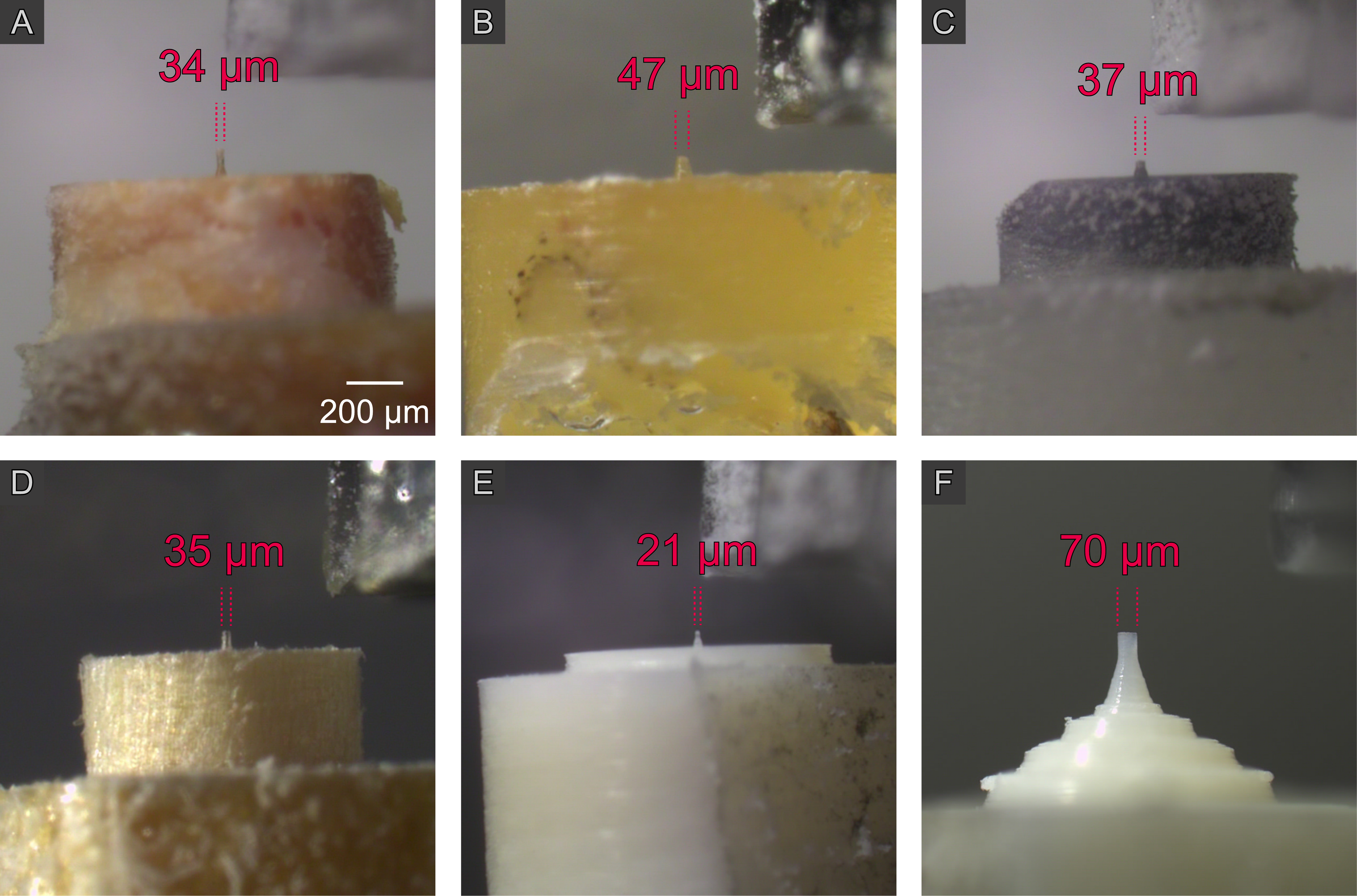}
    \caption{Light microscope images of critical-point-dried sausage (\textbf{A}), paraffin-embedded zebrafish (\textbf{B}), epoxy-embedded moss (\textbf{C}), wooden toothpick (\textbf{D}), and elephant ivory (\textbf{E}). The sample pillars are all cut with the sample milling machine. A part of the rotation tool can be seen on the top right of the images. An ivory pillar with a diameter of around \SI{70}{\micro\meter} is prepared for the investigation of the different scan protocols (\textbf{F}).}
    \label{fig:micro_lathe_samples}
\end{figure}

To demonstrate the capabilities of the presented sample milling machine, stable pillars are cut from a wide range of materials. Figure~\ref{fig:micro_lathe_samples} shows representative pillars prepared from soft tissue, wood, and ivory, processed by different preparation techniques (paraffin embedding, epoxy embedding, and critical point drying). For all tested materials and preparation techniques, pillars with diameters below \SI{50}{\micro\meter} are successfully produced. These results in Figure~\ref{fig:micro_lathe_samples} demonstrate that the sample milling machine can reliably produce pillars from various biological preparations, providing suitable samples for nanotomographic imaging.
The ivory pillar in Figure~\ref{fig:micro_lathe_samples} F is further used to experimentally validate the scan time reduction.

\subsection{Scan Time Reduction}
\label{subsec:scan_time_red}
\begin{figure}[htbp]
    \centering
    \includegraphics[width=\linewidth]{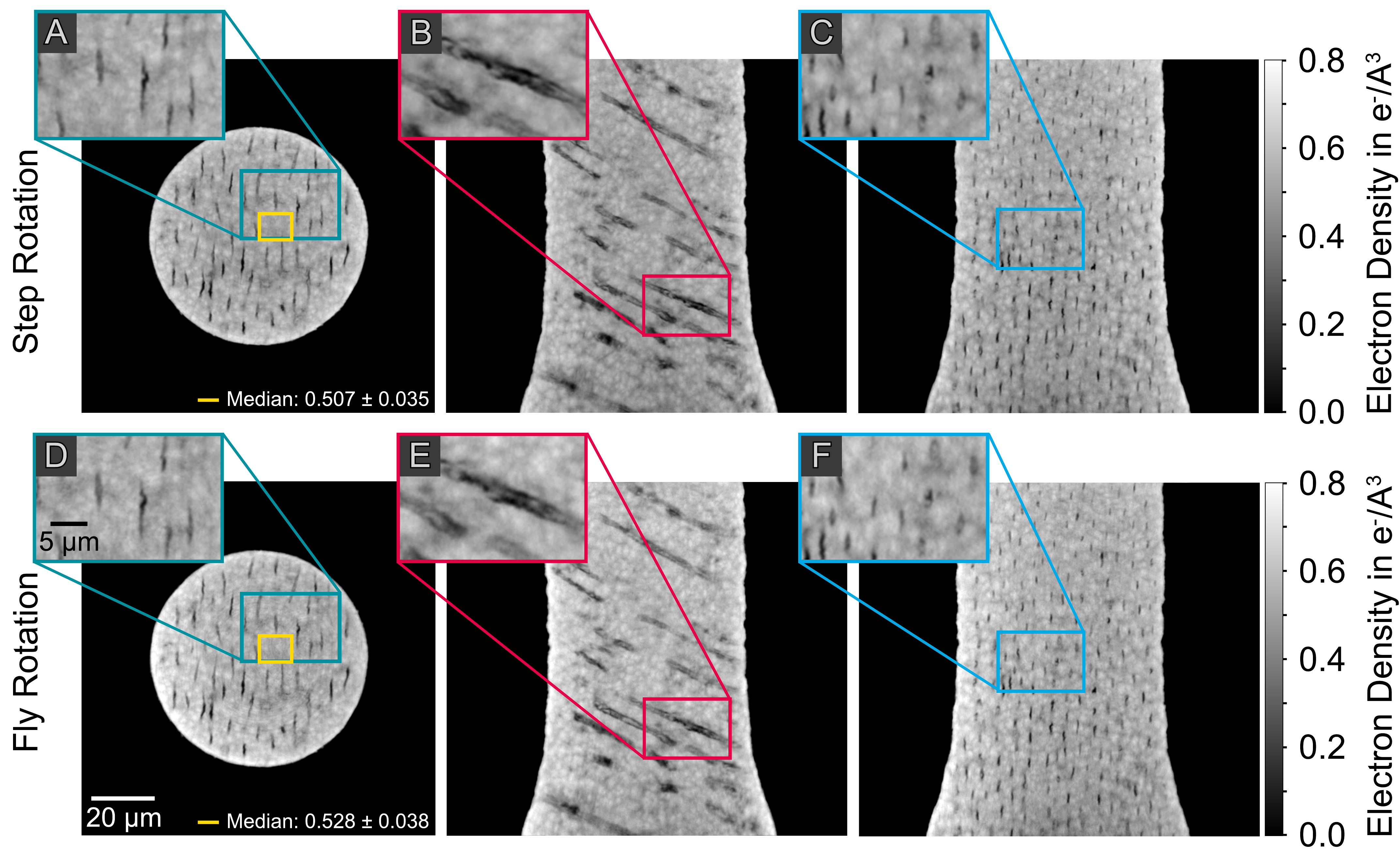}
    \caption{Comparison of the reconstructed electron density of elephant ivory between the step-rotation (\textbf{A}, \textbf{B}, \textbf{C}) and fly-rotation scan (\textbf{D}, \textbf{E}, \textbf{F}). Visually, the two scan methods are nearly indistinguishable. However, the fly-rotation scan shows slight ring artifacts (\textbf{D}) and a small phase offset on the left side (zoomed out regions) in (\textbf{E}, \textbf{F}). Looking at the median electron density in the yellow region of interest in \textbf{A} with $0.507 \pm 0.035$~\si[per-mode=symbol]{\electron\per\cubic\angstrom}, it is slightly lower than the median electron density in \textbf{D} with $0.528 \pm 0.038$~\si[per-mode=symbol]{\electron\per\cubic\angstrom}, but lies within its standard deviation.}
    \label{fig:phase_comp}
\end{figure}

To compare the two scan protocols, an ivory pillar with a diameter of around \SI{70}{\micro\meter} is measured (see Figure~\ref{fig:micro_lathe_samples}~F). Two consecutive scans are performed, starting with a step-rotation scan followed by a fly-rotation scan.
Figure~\ref{fig:phase_comp} compares the retrieved volumes by showing slices through three different axes for both scans. A first visual inspection shows a good agreement between the results of the two scan protocols. Both are capable of revealing fine structures within the ivory (Figure \ref{fig:phase_comp} C and F) while maintaining a consistent electron density. This can be confirmed by calculating the median electron density in the yellow region in Figure~\ref{fig:phase_comp} A and D. The median value for the step-rotation scan of $0.507 \pm 0.035$~\si[per-mode=symbol]{\electron\per\cubic\angstrom} lies within the standard deviation of the median electron density of the fly-rotation scan with $0.528 \pm 0.038$~\si[per-mode=symbol]{\electron\per\cubic\angstrom}.
Both scans achieve very similar resolution values. The step-rotation scan reaches \SI{276}{\nano\meter} and the fly-rotation scan \SI{271}{\nano\meter} in resolution based on the Fourier shell correlation (FSC) half-bit criterion (see Supplements).
In both scans, the tubules and small density changes within the solid parts of the ivory are clearly visible, as observed in other samples before~\cite{Alberic2018, Flenner2020holo}. However, upon close inspection of Figure~\ref{fig:phase_comp} A and D, two differences are discernible.

The first difference is visible in Figure~\ref{fig:phase_comp}~D, where slight ring artifacts are visible, unlike in A. In fact, multiple rings with different centers are present. These centers likely correspond to the different centers of rotation of the tomograms acquired during the fly-rotation scan. Here, it is noteworthy that, despite each phase projection being retrieved from 16 interference patterns with different sample positions, the ring artifacts become visible in the reconstructed slices. Hence, some sort of noise or pixel correlation seems to propagate through the signal retrieval of \textit{PtyPy}, which might be linked to the fly-rotation scan approach. However, the artifacts can be corrected by standard ring-removal algorithms during volume reconstruction.

The second difference is a slight remaining phase gradient in the fly-rotation scan, which can be seen in the bottom left of the non-zoomed image of Figure~\ref{fig:phase_comp} D, when compared to A. This gradient can also be seen in Figure~\ref{fig:phase_comp} E and F on the left side. A possible reason could be an error in the previously run step-rotation scan, which led to a drift in the center of rotation over roughly 120 (binned) pixels in the time range of a full tomogram. Since the fly-rotation scan was started directly afterwards, it would have been performed in a slight off-axis geometry. This would have resulted in a less-ideal stepping pattern shifted towards the edge of the detector for some projection angles, making phase retrieval and ramp correction more challenging. The difference compared to the step-rotation scan might be reduced by increasing the number of reconstruction iterations for the fly-rotation scan. However, here, the number of iterations is kept constant for both scans to maintain the comparability.

Table \ref{tab:scan_param} shows the total scan times for the step-rotation and fly-rotation scan, as well as the actual total exposure time per scan. With the fly-rotation protocol, the scan time can be reduced from \SI{6}{\hour} \SI{4}{\min} to \SI{3}{\hour} \SI{41}{\min}. By subtracting the total exposure time from the total scan time, the overhead time can be calculated. With the fly-rotation scan, the overhead is reduced by a factor of $11$ from \SI{3}{\hour} \SI{20}{\minute} to \SI{18}{\minute}, showing a dramatic increase in scan efficiency while maintaining resolution.

\begin{table}
    \caption{Overview of the experimental parameters for the shown scans. Each projection is retrieved from 16 interference patterns, acquired in a raster scan pattern~\cite{Clare2015}. The step-rotation and fly-rotation scans are conducted at a magnification of $77.60$, and the fly-rotation II scan at $65.21$.}
    \setlength{\tabcolsep}{3pt}
    \centering
    \begin{tabular}{|l|c|c|c|c|c|c|}
        \hline
          & \textbf{resolution} & \textbf{\# of proj.} & \textbf{exp. time} & \textbf{tot. exp. time} & \textbf{tot. scan time} & \textbf{overhead} \\
        \hline
        \textbf{step rotation} & \SI{276}{\nano\meter} & 1532 & \SI{0.5}{\second} & \SI{3}{\hour} \SI{24}{\minute} & \SI{6}{\hour} \SI{04}{\minute} & \SI{3}{\hour} \SI{20}{\minute}\\
        \hline
        \textbf{fly rotation} & \SI{271}{\nano\meter} & 1522 & \SI{0.5}{\second} & \SI{3}{\hour} \SI{23}{\minute} & \SI{3}{\hour} \SI{41}{\minute} & \SI{18}{\minute}\\
        \hline
        \textbf{fly rotation II} & \SI{335}{\nano\meter} & 930 & \SI{0.25}{\second} & \SI{1}{\hour} \SI{2}{\minute} & \SI{1}{\hour} \SI{16}{\minute} & \SI{14}{\minute}\\
        \hline
    \end{tabular}
    \label{tab:scan_param}
\end{table}

\begin{figure}[htbp]
    \centering
    \includegraphics[width=\linewidth]{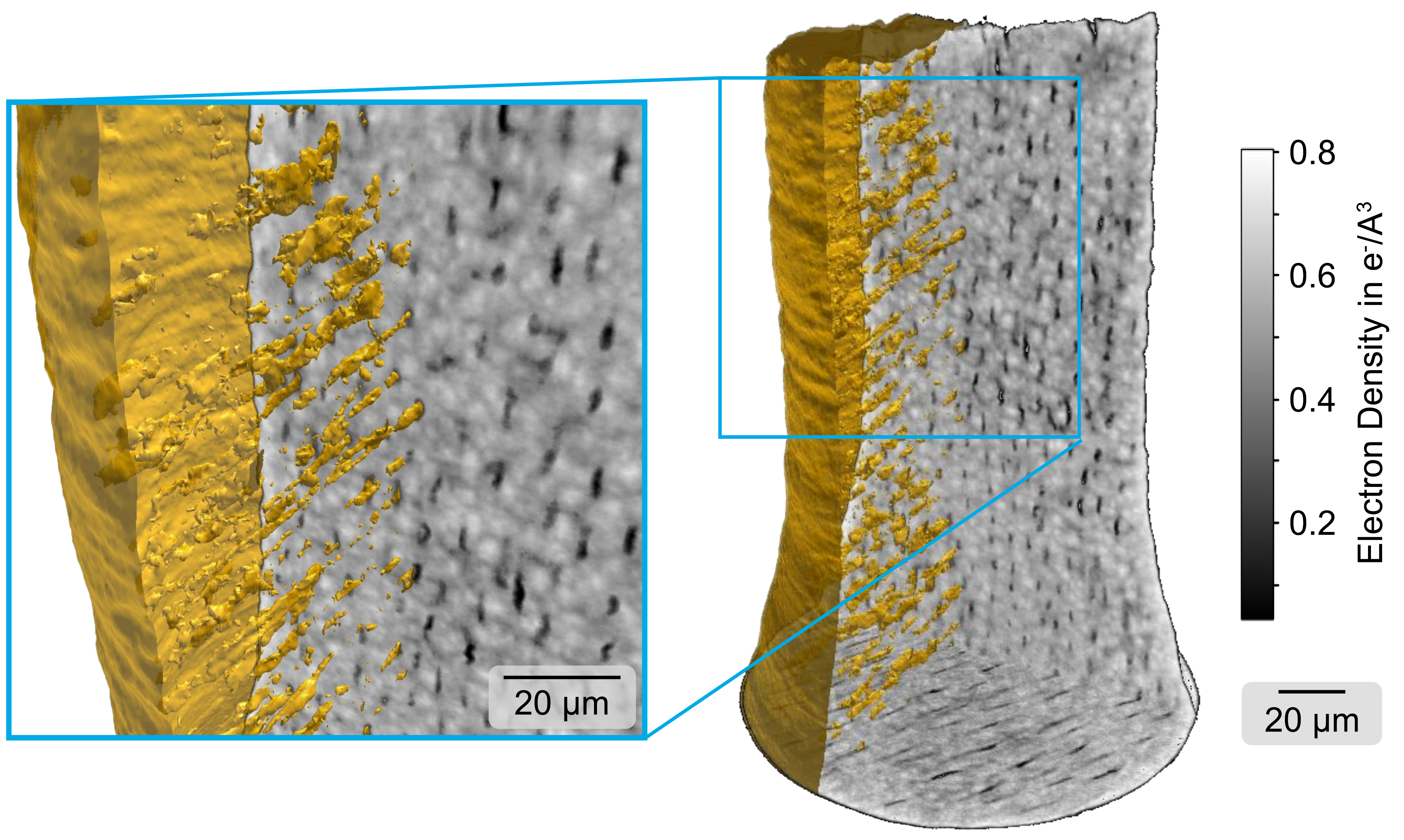}
    \caption{Fast fly-rotation ptychographic tomogram, which was acquired in \SI{72}{\minute} total scan time. Using a simple thresholding operation, the tubules within the elephant tusk can be segmented and visualized as an iso surface. The fly-rotation II scan reaches a resolution of \SI{335}{\nano\meter} based on the half-bit Fourier shell correlation.}
    \label{fig:fast_scan}
\end{figure}

The reconstructed volume of the second fly-rotation scan (fly-rotation II) is shown in Figure~\ref{fig:fast_scan}. By reducing the exposure time and the number of projections, the whole ptychographic tomogram was acquired in \SI{72}{\minute}, with an overhead of \SI{14}{\minute}. While the scan shows slightly more noise than the previous scans, the sample structure is well maintained and allows segmentation of the tubules using a simple thresholding operation. Due to the lower magnification, exposure time, and fewer projections, a resolution of \SI{335}{\nano\meter} is reached (see Supplements).

\section{Discussion}
While near-field ptychography is a powerful full-field imaging technique, it is currently limited by long acquisition times, which reduce the number of scans that can be performed within the limited beam time available at synchrotrons.
This work shows that the overhead in near-field ptychography can be reduced by an order of magnitude through the introduction of continuous sample rotation, thereby addressing its main bottleneck. This brings the total scan times of ptychographic tomograms down to nearly an hour.
While the concept of a continuous sample movement is not new in the field of ptychography, it has only been applied to far-field ptychography. However, in far-field ptychography, the sample is stepped over multiple hundreds or even thousands of positions at each projection angle, and often exceeds the total number of projection angles. In contrast, near-field ptychography can be performed with as few as six interference patterns~\cite{Clare2015}. This makes far-field ptychography more dependent on the stepping speed than the rotation speed. Hence, fly-scan approaches for far-field ptychography introduce a continuous translational movement of the sample to lower the impact of the stepping overhead~\cite{Clark2014, Pelz2014, Huang2015, Odstrcil2018, Deng2019, Jiang2021, Jones2022} and can therefore be separated from the approach of continuous rotation presented here.
In D. Batey et al.~\cite{Batey2022}, a tomogram was acquired in a far-field experiment with 600 projections and $150 \times 20$ steps per projection in \SI{3}{\hour}. A sample volume of around $(\SI{20}{\micro\meter})^3$ was measured with a resolution of \SI{250}{\nano\meter} (FSC on the volume).
In another far-field experiment, S. Cipiccia et al.~\cite{Cipiccia2024} report a tomogram with 501 projections and $1000 \times 100$ steps per projection in around \SI{7}{\hour}, in which a volume of around $(\SI{100}{\micro\meter})^3$ has been scanned with a resolution of \SI{270}{\nano\meter} (Fourier ring correlation on slices).
To set this work in context, the fly-rotation tomogram of the elephant ivory was measured with 1522 projections and $4 \times 4$ steps per projection in less than \SI{4}{\hour} with a volume of $(\SI{130}{\micro\meter})^3$ and a resolution of \SI{271}{\nano\meter} (binned).
By reducing the exposure time to \SI{0.25}{\second} and the number of projections to 930, the second fly-rotation scan was performed in \SI{72}{\minute}, measuring a volume of $(\SI{169}{\micro\meter})^3$ with a resolution of \SI{335}{\nano\meter} (binned), even though PETRA III was running in the less optimal 40-bunch mode with a reduced current of \SI{110}{\milli\ampere}. This makes the presented setup the fastest ptychographic nanotomography setup currently available.

In these experiments, an FZP is used as focusing optics since the same setup is also utilized for full-field holotomography, which requires a smooth, speckle-free wavefront \cite{Flenner2020holo}. However, for near-field ptychography, the variation in the wave front is desired. Therefore, near-field ptychography comes with fewer optics requirements and allows the use of a broader range of focusing optics. One example is diamond lenses \cite{Snigireva2021, Wang2025}. While they are currently not yet able to completely suppress the appearance of speckle patterns at \SI{11}{\kilo\electronvolt}, they drastically increase the flux. At P05, the use of diamond lenses boosts the flux by a factor of four compared to an FZP. This allows reducing the exposure time by a factor of four, while maintaining the same photon statistics. For a full ptychographic tomography fly-rotation scan, the total scan time could likely be reduced to \SI{30}{\minute}, while a step-rotation scan would still need around \SI{4}{\hour}. In combination with the planned update to PETRA IV, which will drastically improve flux, this will result in ultra-short scan times and might even enable in-situ experiments. Furthermore, an increase in flux and a reduction in exposure time will relax the stability constraint, which is especially critical for the fly-rotation scan, and will make it less prone to signal retrieval problems due to probe or sample changes.

Sandpaper is often used as a modulator since it is cheap, abundant, and has a desirable, strong variation in the wavefront. However, during long scan times, a noticeable drift occurs. To have a strong overlap constraint in near-field ptychography, it is important to have a stable probe during the acquisition of a set of interference patterns. Otherwise, the quality of the retrieved signals will significantly degrade, since a modulator drift could be interpreted as a sample shift during the signal retrieval.
A modulator drift is not a concern as long as the drift is sufficiently small during the acquisition time of a full set of interference patterns. For a step-rotation scan, the required time for a set of interference patterns is fairly small since the interference patterns are acquired in succession. However, in the fly-rotation scan, a whole tomogram is performed between two interference patterns, increasing the stability constraint of the modulator. Since all sets are only complete after the last tomogram is performed, the modulator must be stable during the whole scan time (nearly 4h in this example). 
As standard sandpaper proved insufficient for these requirements, alternative materials were investigated. Diamond plates were found to provide the required long-term stability and were therefore used in the experiments shown.

The diamond plate is not only a stable wave front marker, but since it can be manufactured precisely by laser ablation, it can also be used to further study the impact of different structure sizes and patterns on the signal retrieval. This allows the systematic optimization of the modulator for different sample types and geometries across different setups. Furthermore, laser ablation is a reproducible production method, allowing for duplicating wavefront markers or making the design widely available. However, with costs in the range of $1000$ EUR, diamond plates are substantially more expensive than sandpaper, but are still way less expensive and more versatile than perturbed Fresnel zone plates~\cite{Odstrcil2019}.
As a proof of example, this work used a diamond plate with a thickness of \SI{500}{\micro\meter}. However, it absorbs around 25\% at \SI{11}{\kilo\electronvolt}. For future experiments, a thinner version of the diamond plate should therefore be used. A thickness of around \SI{30}{\micro\meter} seems feasible and would reduce the absorption to less than 2\%, allowing shorter exposure times while maintaining the statistics. Furthermore, the number of indentations should be increased to amplify wavefront variations, thereby simplifying the signal retrieval.

\section{Conclusion}
In this work, we address the main bottlenecks of quantitative nanotomography with near-field ptychography to enhance its accessibility and usability. By offering a setup with high flexibility and a broad range of magnifications, we open up the possibility to scan samples of various types and sizes. Based on this setup, we were able to drastically speed up the scanning process of near-field ptychography by introducing continuous sample rotation during image acquisition, enabling the high throughput required for statistical analysis of complex systems. This makes the proposed setup the fastest ptychography setup currently available. To overcome the limitations of FIB preparation for biological samples, we additionally presented a mechanical sample milling machine that can prepare samples with various sizes and materials down to \SI{20}{\micro\meter} in diameter.

\begin{backmatter}

\bmsection{Funding} The authors gratefully acknowledge financial support by the ERC consolidator grant (Julia Herzen, TUM, DEPICT, PE3, 101125761) and by the EIC pathfinder (1MICRON, 101186826).

\bmsection{Disclosures} The authors declare no conflicts of interest.

\bmsection{Data availability} Data underlying the results presented in this paper are not publicly available at this time but may be obtained from the authors upon reasonable request.

\bmsection{Acknowledgments}
We acknowledge Helmholtz-Zentrum Hereon (Geesthacht, Germany) and DESY (Hamburg, Germany) for providing the experimental facilities. This research was carried out at PETRA~III. Beamtime was allocated for proposal I-20221419 and I-20240219. Furthermore, we thank Martin Dierolf and Benedikt Günther for their support and ideas during the first measurements. We thank Pierre Thibault, Björn Enders, and Benedikt Daurer for organizing the \textit{PtyPy} workshops with many fruitful discussions and very helpful tutorials.
We also thank Julio Cesar da Silva for helping us with the implementation of toupy and Ana Diaz for her useful tips and recommendations regarding near-field ptychography.
Many thanks to Jörg Hammel and Madleen Busse for providing interesting samples of different preparation techniques to test the capabilities of the sample milling machine.
This research was supported through the Maxwell computational resources operated at Deutsches Elektronen-Synchrotron DESY, Hamburg, Germany.

\bmsection{Author contributions}
S.W. wrote the main manuscript.
S.W., S.F., I.G., D.J., S.S., and J.He. conceptualized the manuscript.
S.W., S.F., I.G., S.B., F.H., M.W., and J.Ha. conducted the experiments.
S.W., J.B., and H.B. conceptualized the sample lathe.
J.B. and H.B. built the sample lathe.
S.W., S.B., and B.D. evaluated the data.
J.Ha. and F.S. constructed and created the diamond modulator.
K.S. added valuable insights into FIB sample preparation.
F.V. provided the elephant ivory and specimen related insights.
All authors reviewed the manuscript.

\end{backmatter}


\bibliography{reference}

@article{Holler2020,
author = {Holler, Mirko and Ihli, Johannes and Tsai, Esther H. R. and Nudelman, Fabio and Verezhak, Mariana and van de Berg, Wilma D. J. and Shahmoradian, Sarah H.},
doi = {10.1107/S1600577519017028},
issn = {1600-5775},
journal = {Journal of Synchrotron Radiation},
month = {3},
number = {2},
pages = {472--476},
pmid = {32153287},
title = {{A lathe system for micrometre-sized cylindrical sample preparation at room and cryogenic temperatures}},
url = {https://scripts.iucr.org/cgi-bin/paper?S1600577519017028},
volume = {27},
year = {2020}
}

@phdthesis{Messler2021,
author = {Messler, Olivia},
school = {KTH Royal Institute of Technology},
title = {{Development of a CNC Milling System for Preparation of Micrometer-Sized Samples for X-ray Nanotomography}},
year = {2021}
}

@article{Flenner2020holo,
author = {Flenner, Silja and Kubec, Adam and David, Christian and Storm, Malte and Schaber, Clemens F. and Vollrath, Fritz and M{\"{u}}ller, Martin and Greving, Imke and Hagemann, Johannes},
doi = {10.1364/OE.406074},
issn = {1094-4087},
journal = {Optics Express},
month = {12},
number = {25},
pages = {37514},
pmid = {33379584},
title = {{Hard X-ray nano-holotomography with a Fresnel zone plate}},
url = {https://opg.optica.org/abstract.cfm?URI=oe-28-25-37514},
volume = {28},
year = {2020}
}

@article{Flenner2020,
author = {Flenner, Silja and Storm, Malte and Kubec, Adam and Longo, Elena and Doring, Florian and Pelt, Daniel M. and David, Christian and Muller, Martin and Greving, Imke},
doi = {10.1107/S1600577520007407},
isbn = {1600577520},
issn = {16005775},
journal = {Journal of Synchrotron Radiation},
pages = {1339--1346},
pmid = {32876609},
title = {{Pushing the temporal resolution in absorption and Zernike phase contrast nanotomography: Enabling fast in situ experiments}},
volume = {27},
year = {2020}
}

@inproceedings{Flenner2022p,
author = {Flenner, Silja and Hagemann, Johannes and Storm, Malte and Kubec, Adam and Qi, Peng and David, Christian and Longo, Elena and Niese, Sven and Gawlitza, Peter and Zeller-Plumhoff, Berit and Reimers, Jan and M{\"{u}}ller, Martin and Greving, Imke},
booktitle = {Developments in X-Ray Tomography XIV},
doi = {10.1117/12.2632706},
editor = {M{\"{u}}ller, Bert and Wang, Ge},
isbn = {9781510654686},
month = {11},
number = {November 2022},
pages = {19},
publisher = {SPIE},
title = {{Hard x-ray nanotomography at the P05 imaging beamline at PETRA III}},
url = {https://www.spiedigitallibrary.org/conference-proceedings-of-spie/12242/2632706/Hard-x-ray-nanotomography-at-the-P05-imaging-beamline-at/10.1117/12.2632706.full},
year = {2022}
}

@article{Wirtensohn2024,
author = {Wirtensohn, Sami and Qi, Peng and David, Christian and Herzen, Julia and Greving, Imke and Flenner, Silja},
doi = {10.1364/OPTICA.524812},
issn = {2334-2536},
journal = {Optica},
month = {6},
number = {6},
pages = {852--859},
title = {{Nanoscale dark-field imaging in full-field transmission X-ray microscopy}},
url = {https://opg.optica.org/abstract.cfm?URI=optica-11-6-852},
volume = {11},
year = {2024}
}

@article{Wirtensohn2025,
author = {Sami Wirtensohn and Silja Flenner and Dominik John and Peng Qi and Christian David and Manfred May and Patrick Huber and Dirk Herzog and Stefan Tangl and Carina Kampleitner and Kritika Singh and Ingomar Kelbassa and Katrin Bekes and Julia Herzen and Imke Greving},
doi = {10.1038/s41377-026-02263-z},
issn = {2047-7538},
number = {1},
journal = {Light: Science \& Applications},
month = {5},
pages = {223},
title = {Directional dark field for nanoscale full-field transmission X-ray microscopy},
volume = {15},
url = {https://www.nature.com/articles/s41377-026-02263-z http://arxiv.org/abs/2506.16998},
year = {2026}
}

@article{John2024,
author = {John, Dominik and Gottwald, Wolfgang and Berthe, Daniel and Wirtensohn, Sami and Hickler, Julia and Heck, Lisa and Herzen, Julia},
doi = {10.1038/s41598-024-56201-3},
isbn = {0123456789},
issn = {2045-2322},
journal = {Scientific Reports},
month = {3},
number = {1},
pages = {5599},
publisher = {Nature Publishing Group UK},
title = {{X-ray dark-field computed tomography for monitoring of tnumber freezing}},
url = {https://doi.org/10.1038/s41598-024-56201-3 https://www.nature.com/articles/s41598-024-56201-3},
volume = {14},
year = {2024}
}

@article{Gursoy2014,
author = {G{\"{u}}rsoy, Dogˇa and {De Carlo}, Francesco and Xiao, Xianghui and Jacobsen, Chris},
doi = {10.1107/S1600577514013939},
issn = {1600-5775},
journal = {Journal of Synchrotron Radiation},
month = {9},
number = {5},
pages = {1188--1193},
pmid = {25178011},
title = {{TomoPy: a framework for the analysis of synchrotron tomographic data}},
url = {https://journals.iucr.org/paper?S1600577514013939},
volume = {21},
year = {2014}
}

@article{Pelt2016,
author = {Pelt, Dani{\"{e}}l M. and G{\"{u}}rsoy, Dogˇa and Palenstijn, Willem Jan and Sijbers, Jan and {De Carlo}, Francesco and Batenburg, Kees Joost},
doi = {10.1107/S1600577516005658},
issn = {1600-5775},
journal = {Journal of Synchrotron Radiation},
month = {5},
number = {3},
pages = {842--849},
title = {{Integration of TomoPy and the ASTRA toolbox for advanced processing and reconstruction of tomographic synchrotron data}},
url = {https://journals.iucr.org/paper?S1600577516005658},
volume = {23},
year = {2016}
}

@article{Guizar-Sicairos2011,
author = {Guizar-Sicairos, Manuel and Diaz, Ana and Holler, Mirko and Lucas, Miriam S. and Menzel, Andreas and Wepf, Roger A. and Bunk, Oliver},
doi = {10.1364/OE.19.021345},
issn = {1094-4087},
journal = {Optics Express},
month = {10},
number = {22},
pages = {21345},
pmid = {22108985},
title = {{Phase tomography from x-ray coherent diffractive imaging projections}},
url = {https://opg.optica.org/abstract.cfm?URI=oe-19-22-21345},
volume = {19},
year = {2011}
}

@inproceedings{daSilvia2017,
author = {da Silva, Julio C. and Haubrich, Jan and Requena, Guillermo and Hubert, Maxime and Pacureanu, Alexandra and Bloch, Leonid and Yang, Yang and Cloetens, Peter},
booktitle = {Developments in X-Ray Tomography XI},
doi = {10.1117/12.2272971},
editor = {M{\"{u}}ller, Bert and Wang, Ge},
isbn = {9781510612396},
issn = {1996756X},
month = {9},
pages = {5},
publisher = {SPIE},
title = {{High energy near- and far-field ptychographic tomography at the ESRF}},
url = {https://www.spiedigitallibrary.org/conference-proceedings-of-spie/10391/2272971/High-energy-near--and-far-field-ptychographic-tomography-at/10.1117/12.2272971.full},
volume = {10391},
year = {2017}
}

@article{Enders2016,
author = {Enders, B. and Thibault, P.},
doi = {10.1098/rspa.2016.0640},
issn = {1364-5021},
journal = {Proceedings of the Royal Society A: Mathematical, Physical and Engineering Sciences},
month = {12},
number = {2196},
pages = {20160640},
title = {{A computational framework for ptychographic reconstructions}},
url = {https://royalsocietypublishing.org/doi/10.1098/rspa.2016.0640},
volume = {472},
year = {2016}
}

@article{Thibault2008,
author = {Thibault, Pierre and Dierolf, Martin and Menzel, Andreas and Bunk, Oliver and David, Christian and Pfeiffer, Franz},
doi = {10.1126/science.1158573},
issn = {0036-8075},
journal = {Science},
month = {7},
number = {5887},
pages = {379--382},
title = {{High-Resolution Scanning X-ray Diffraction Microscopy}},
url = {https://www.science.org/doi/10.1126/science.1158573},
volume = {321},
year = {2008}
}

@article{Giewekemeyer2010,
author = {Giewekemeyer, Klaus and Thibault, Pierre and Kalbfleisch, Sebastian and Beerlink, Andr{\'{e}} and Kewish, Cameron M. and Dierolf, Martin and Pfeiffer, Franz and Salditt, Tim},
doi = {10.1073/pnas.0905846107},
issn = {0027-8424},
journal = {Proceedings of the National Academy of Sciences},
month = {1},
number = {2},
pages = {529--534},
pmid = {20018650},
title = {{Quantitative biological imaging by ptychographic x-ray diffraction microscopy}},
url = {http://www.ncbi.nlm.nih.gov/pubmed/20018650 http://www.pubmedcentral.nih.gov/articlerender.fcgi?artid=PMC2795774 https://pnas.org/doi/full/10.1073/pnas.0905846107},
volume = {107},
year = {2010}
}

@article{Birnbacher2021,
author = {Birnbacher, Lorenz and Braig, Eva-Maria and Pfeiffer, Daniela and Pfeiffer, Franz and Herzen, Julia},
doi = {10.1007/s00259-021-05259-6},
isbn = {0025902105},
issn = {1619-7070},
journal = {European Journal of Nuclear Medicine and Molecular Imaging},
month = {12},
number = {13},
pages = {4171--4188},
pmid = {33846846},
publisher = {European Journal of Nuclear Medicine and Molecular Imaging},
title = {{Quantitative X-ray phase contrast computed tomography with grating interferometry}},
url = {https://link.springer.com/10.1007/s00259-021-05259-6},
volume = {48},
year = {2021}
}

@article{Wang2025,
author = {Wang, Wenxin and D{\"{o}}hrmann, Ralph and Botta, Stephan and Madsen, Anders and Schroer, Christian G. and Seiboth, Frank},
doi = {10.1364/OE.562556},
issn = {1094-4087},
journal = {Optics Express},
month = {6},
number = {11},
pages = {22349},
pmid = {40515226},
title = {{Diamond X-ray lens cubes with integrated aberration compensation}},
url = {https://opg.optica.org/abstract.cfm?URI=oe-33-11-22349},
volume = {33},
year = {2025}
}

@article{Snigireva2021,
author = {Snigireva, Irina and Polikarpov, Maxim and Snigirev, Anatoly},
doi = {10.1080/08940886.2021.2022387},
issn = {0894-0886},
journal = {Synchrotron Radiation News},
month = {11},
number = {6},
pages = {12--20},
publisher = {Taylor and Francis},
title = {{Diamond X-Ray Refractive Optics}},
url = {https://doi.org/10.1080/08940886.2021.2022387 https://www.tandfonline.com/doi/full/10.1080/08940886.2021.2022387},
volume = {34},
year = {2021}
}

@article{Wang2019,
author = {Wang, Hongchang and Atwood, Robert C. and Pankhurst, Matthew James and Kashyap, Yogesh and Cai, Biao and Zhou, Tunhe and Lee, Peter David and Drakopoulos, Michael and Sawhney, Kawal},
doi = {10.1038/s41598-019-45561-w},
issn = {2045-2322},
journal = {Scientific Reports},
month = {6},
number = {1},
pages = {8913},
pmid = {31222085},
title = {{High-energy, high-resolution, fly-scan X-ray phase tomography}},
url = {https://www.nature.com/articles/s41598-019-45561-w},
volume = {9},
year = {2019}
}

@article{Schaff2020,
archivePrefix = {arXiv},
arxivId = {1910.04309},
author = {Schaff, Florian and Morgan, Kaye S. and Pollock, James A. and Croton, Linda C. P. and Hooper, Stuart B. and Kitchen, Marcus J.},
doi = {10.1109/TMI.2020.3006815},
eprint = {1910.04309},
issn = {0278-0062},
journal = {IEEE Transactions on Medical Imaging},
month = {12},
number = {12},
pages = {3891--3899},
pmid = {32746132},
title = {{Material Decomposition Using Spectral Propagation-Based Phase-Contrast X-Ray Imaging}},
url = {https://ieeexplore.ieee.org/document/9133129/},
volume = {39},
year = {2020}
}

@article{Taphorn2022,
author = {Taphorn, Kirsten and Busse, Madleen and Brantl, Johannes and G{\"{u}}nther, Benedikt and Diaz, Ana and Holler, Mirko and Dierolf, Martin and Mayr, Doris and Pfeiffer, Franz and Herzen, Julia},
doi = {10.1002/advs.202201723},
issn = {21983844},
journal = {Advanced Science},
number = {24},
pages = {1--13},
pmid = {35748171},
title = {{X-ray Stain Localization with Near-Field Ptychographic Computed Tomography}},
volume = {9},
year = {2022}
}

@article{RajaSomu2025,
author = {{Raja Somu}, Dawn and Soini, Steven A. and Briggs, Ani and Singh, Kritika and Greving, Imke and Porter, Marianne and Passerotti, Michelle and Merk, Vivian},
doi = {10.1021/acsnano.5c02004},
issn = {1936086X},
journal = {ACS Nano},
number = {14},
pages = {14410--14421},
pmid = {40191917},
title = {{A Nanoscale View of the Structure and Deformation Mechanism of Mineralized Shark Vertebral Cartilage}},
volume = {19},
year = {2025}
}

@article{David2002,
author = {David, C. and N{\"{o}}hammer, B. and Solak, H. H. and Ziegler, E.},
doi = {10.1063/1.1516611},
issn = {0003-6951},
journal = {Applied Physics Letters},
month = {10},
number = {17},
pages = {3287--3289},
title = {{Differential x-ray phase contrast imaging using a shearing interferometer}},
url = {https://pubs.aip.org/apl/article/81/17/3287/115118/Differential-x-ray-phase-contrast-imaging-using-a},
volume = {81},
year = {2002}
}

@article{Berthe2024,
author = {Berthe, Daniel and Heck, Lisa and Resch, Sandra and Dierolf, Martin and Brantl, Johannes and G{\"{u}}nther, Benedikt and Petrich, Christian and Achterhold, Klaus and Pfeiffer, Franz and Grandl, Susanne and Hellerhoff, Karin and Herzen, Julia},
doi = {10.1038/s41598-024-77346-1},
isbn = {4159802477346},
issn = {2045-2322},
journal = {Scientific Reports},
month = {10},
number = {1},
pages = {25576},
pmid = {39462058},
title = {{Grating-based phase-contrast computed tomography for breast tnumber at an inverse compton source}},
url = {https://www.nature.com/articles/s41598-024-77346-1},
volume = {14},
year = {2024}
}

@article{Fitzgerald2000,
author = {Fitzgerald, Richard},
doi = {10.1063/1.1292471},
issn = {0031-9228},
journal = {Physics Today},
month = {7},
number = {7},
pages = {23--26},
title = {{Phase‐Sensitive X‐Ray Imaging}},
url = {http://physicstoday.scitation.org/doi/10.1063/1.1292471},
volume = {53},
year = {2000}
}

@article{Alberic2018,
author = {Alb{\'{e}}ric, M. and Gourrier, A. and Wagermaier, W. and Fratzl, P. and Reiche, I.},
doi = {10.1016/j.actbio.2018.02.016},
issn = {17427061},
journal = {Acta Biomaterialia},
month = {5},
pages = {342--351},
pmid = {29477454},
title = {{The three-dimensional arrangement of the mineralized collagen fibers in elephant ivory and its relation to mechanical and optical properties}},
url = {https://linkinghub.elsevier.com/retrieve/pii/S1742706118300862},
volume = {72},
year = {2018}
}

@article{Lioliou2024,
author = {Lioliou, Grammatiki and {Roche i Morg{\'{o}}}, Oriol and Astolfo, Alberto and Zekavat, Amir Reza and Endrizzi, Marco and Bate, David and Cipiccia, Silvia and Olivo, Alessandro and Hagen, Charlotte},
doi = {10.1016/j.tmater.2024.100034},
issn = {2949673X},
journal = {Tomography of Materials and Structures},
month = {6},
number = {November 2023},
pages = {100034},
title = {{Recent developments in fly scan methods for phase and multi-contrast x-ray micro-CT based on amplitude modulated beams}},
url = {https://linkinghub.elsevier.com/retrieve/pii/S2949673X24000111},
volume = {5},
year = {2024}
}

@article{Stockmar2013,
author = {Stockmar, Marco and Cloetens, Peter and Zanette, Irene and Enders, Bjoern and Dierolf, Martin and Pfeiffer, Franz and Thibault, Pierre},
doi = {10.1038/srep01927},
issn = {2045-2322},
journal = {Scientific Reports},
month = {5},
number = {1},
pages = {1927},
publisher = {Nature Publishing Group},
title = {{Near-field ptychography: phase retrieval for inline holography using a structured illumination}},
url = {http://www.nature.com/articles/srep01927 https://www.nature.com/articles/srep01927},
volume = {3},
year = {2013}
}

@article{Shirani2024,
author = {Shirani, Shiva and Cuesta, Ana and Santacruz, Isabel and {De la Torre}, Angeles G. and Diaz, Ana and Trtik, Pavel and Holler, Mirko and Aranda, Miguel A.G.},
doi = {10.1016/j.cemconres.2024.107622},
issn = {00088846},
journal = {Cement and Concrete Research},
number = {4},
pages = {107622},
publisher = {Elsevier Ltd},
title = {{X-ray near-field ptychographic nanoimaging of cement pastes}},
url = {https://doi.org/10.1016/j.cemconres.2024.107622},
volume = {185},
year = {2024}
}

@article{Dierolf2010,
author = {Dierolf, Martin and Menzel, Andreas and Thibault, Pierre and Schneider, Philipp and Kewish, Cameron M. and Wepf, Roger and Bunk, Oliver and Pfeiffer, Franz},
doi = {10.1038/nature09419},
issn = {0028-0836},
journal = {Nature},
month = {9},
number = {7314},
pages = {436--439},
pmid = {20864997},
publisher = {Nature Publishing Group},
title = {{Ptychographic X-ray computed tomography at the nanoscale}},
url = {https://www.nature.com/articles/nature09419},
volume = {467},
year = {2010}
}

@article{Clare2015,
author = {Clare, Richard M. and Stockmar, Marco and Dierolf, Martin and Zanette, Irene and Pfeiffer, Franz},
doi = {10.1364/OE.23.019728},
issn = {1094-4087},
journal = {Optics Express},
month = {7},
number = {15},
pages = {19728},
title = {{Characterization of near-field ptychography}},
url = {https://opg.optica.org/abstract.cfm?URI=oe-23-15-19728},
volume = {23},
year = {2015}
}

@article{Cipiccia2024,
author = {Cipiccia, Silvia and Fratini, Michela and Erin, Ecem and Palombo, Marco and Vogel, Silvia and Burian, Max and Zhou, Fenglei and Parker, Geoff J. M. and Batey, Darren J.},
doi = {10.1140/epjp/s13360-024-05224-w},
isbn = {0123456789},
issn = {2190-5444},
journal = {The European Physical Journal Plus},
month = {5},
number = {5},
pages = {434},
publisher = {Springer Berlin Heidelberg},
title = {{Fast X-ray ptychography: towards nanoscale imaging of large volume of brain}},
url = {https://doi.org/10.1140/epjp/s13360-024-05224-w https://link.springer.com/10.1140/epjp/s13360-024-05224-w},
volume = {139},
year = {2024}
}

@article{Batey2022,
author = {Batey, Darren and Rau, Christoph and Cipiccia, Silvia},
doi = {10.1038/s41598-022-11292-8},
isbn = {4159802211292},
issn = {2045-2322},
journal = {Scientific Reports},
month = {5},
number = {1},
pages = {7846},
pmid = {35551474},
publisher = {Nature Publishing Group UK},
title = {{High-speed X-ray ptychographic tomography}},
url = {https://doi.org/10.1038/s41598-022-11292-8 https://www.nature.com/articles/s41598-022-11292-8},
volume = {12},
year = {2022}
}

@article{Odstrcil2018,
author = {Odstr{\v{c}}il, Michal and Holler, Mirko and Guizar-Sicairos, Manuel},
doi = {10.1364/OE.26.012585},
issn = {1094-4087},
journal = {Optics Express},
month = {5},
number = {10},
pages = {12585},
pmid = {29801297},
title = {{Arbitrary-path fly-scan ptychography}},
url = {https://opg.optica.org/abstract.cfm?URI=oe-26-10-12585},
volume = {26},
year = {2018}
}

@article{Deng2019,
author = {Deng, Junjing and Preissner, Curt and Klug, Jeffrey A. and Mashrafi, Sheikh and Roehrig, Christian and Jiang, Yi and Yao, Yudong and Wojcik, Michael and Wyman, Max D. and Vine, David and Yue, Ke and Chen, Si and Mooney, Tim and Wang, Maoyu and Feng, Zhenxing and Jin, Dafei and Cai, Zhonghou and Lai, Barry and Vogt, Stefan},
doi = {10.1063/1.5103173},
issn = {0034-6748},
journal = {Review of Scientific Instruments},
month = {8},
number = {8},
pmid = {31472643},
publisher = {AIP Publishing, LLC},
title = {{The Velociprobe: An ultrafast hard X-ray nanoprobe for high-resolution ptychographic imaging}},
url = {https://doi.org/10.1063/1.5103173 https://pubs.aip.org/rsi/article/90/8/083701/360246/The-Velociprobe-An-ultrafast-hard-X-ray-nanoprobe},
volume = {90},
year = {2019}
}

@article{Huang2015,
author = {Huang, Xiaojing and Lauer, Kenneth and Clark, Jesse N. and Xu, Weihe and Nazaretski, Evgeny and Harder, Ross and Robinson, Ian K. and Chu, Yong S.},
doi = {10.1038/srep09074},
issn = {2045-2322},
journal = {Scientific Reports},
month = {3},
number = {1},
pages = {9074},
title = {{Fly-scan ptychography}},
url = {https://www.nature.com/articles/srep09074},
volume = {5},
year = {2015}
}

@article{Jiang2021,
author = {Jiang, Yi and Deng, Junjing and Yao, Yudong and Klug, Jeffrey A. and Mashrafi, Sheikh and Roehrig, Christian and Preissner, Curt and Marin, Fabricio S. and Cai, Zhonghou and Lai, Barry and Vogt, Stefan},
doi = {10.1063/5.0067197},
issn = {0003-6951},
journal = {Applied Physics Letters},
month = {9},
number = {12},
publisher = {AIP Publishing LLC},
title = {{Achieving high spatial resolution in a large field-of-view using lensless x-ray imaging}},
url = {https://pubs.aip.org/apl/article/119/12/124101/40322/Achieving-high-spatial-resolution-in-a-large-field},
volume = {119},
year = {2021}
}

@article{Pelz2014,
author = {Pelz, Philipp M. and Guizar-Sicairos, Manuel and Thibault, Pierre and Johnson, Ian and Holler, Mirko and Menzel, Andreas},
doi = {10.1063/1.4904943},
issn = {0003-6951},
journal = {Applied Physics Letters},
month = {12},
number = {25},
title = {{On-the-fly scans for X-ray ptychography}},
url = {http://dx.doi.org/10.1063/1.4904943 https://pubs.aip.org/apl/article/105/25/251101/26515/On-the-fly-scans-for-X-ray-ptychography},
volume = {105},
year = {2014}
}

@article{Clark2014,
author = {Clark, Jesse N. and Huang, Xiaojing and Harder, Ross J. and Robinson, Ian K.},
doi = {10.1364/OL.39.006066},
issn = {0146-9592},
journal = {Optics Letters},
month = {10},
number = {20},
pages = {6066},
pmid = {25361157},
title = {{Continuous scanning mode for ptychography}},
url = {https://opg.optica.org/abstract.cfm?URI=ol-39-20-6066},
volume = {39},
year = {2014}
}

@article{Jones2022,
author = {Jones, Michael W. M. and van Riessen, Grant A. and Phillips, Nicholas W. and Schrank, Christoph E. and Hinsley, Gerard N. and Afshar, Nader and Reinhardt, Juliane and de Jonge, Martin D. and Kewish, Cameron M.},
doi = {10.1107/S1600577521012856},
issn = {1600-5775},
journal = {Journal of Synchrotron Radiation},
month = {3},
number = {2},
pages = {480--487},
pmid = {35254312},
publisher = {International Union of Crystallography},
title = {{High-speed free-run ptychography at the Australian Synchrotron}},
url = {https://journals.iucr.org/paper?S1600577521012856},
volume = {29},
year = {2022}
}

@article{Hu2023,
author = {Hu, Ziyang and Zhang, Yiqian and Li, Peng and Batey, Darren and Maiden, Andrew},
doi = {10.1364/OE.487002},
issn = {1094-4087},
journal = {Optics Express},
number = {10},
pages = {15791},
pmid = {37157672},
title = {{Near-field multi-slice ptychography: quantitative phase imaging of optically thick samples with visible light and X-rays}},
url = {https://opg.optica.org/abstract.cfm?URI=oe-31-10-15791},
volume = {31},
year = {2023}
}

@article{Tsai2016,
author = {Tsai, Esther H. R. and Usov, Ivan and Diaz, Ana and Menzel, Andreas and Guizar-Sicairos, Manuel},
doi = {10.1364/OE.24.029089},
issn = {1094-4087},
journal = {Optics Express},
month = {12},
number = {25},
pages = {29089},
title = {{X-ray ptychography with extended depth of field}},
url = {https://opg.optica.org/abstract.cfm?URI=oe-24-25-29089},
volume = {24},
year = {2016}
}

@article{Holler2018,
author = {Holler, M. and Raabe, J. and Diaz, A. and Guizar-Sicairos, M. and Wepf, R. and Odstrcil, M. and Shaik, F. R. and Panneels, V. and Menzel, A. and Sarafimov, B. and Maag, S. and Wang, X. and Thominet, V. and Walther, H. and Lachat, T. and Vitins, M. and Bunk, O.},
doi = {10.1063/1.5020247},
issn = {0034-6748},
journal = {Review of Scientific Instruments},
month = {4},
number = {4},
pmid = {29716370},
title = {{OMNY—A tOMography Nano crYo stage}},
url = {http://dx.doi.org/10.1063/1.5020247 https://pubs.aip.org/rsi/article/89/4/043706/362160/OMNY-A-tOMography-Nano-crYo-stage},
volume = {89},
year = {2018}
}

@article{Shahmoradian2017,
author = {Shahmoradian, S. H. and Tsai, E. H. R. and Diaz, A. and Guizar-Sicairos, M. and Raabe, J. and Spycher, L. and Britschgi, M. and Ruf, A. and Stahlberg, H. and Holler, M.},
doi = {10.1038/s41598-017-05587-4},
isbn = {4159801705587},
issn = {2045-2322},
journal = {Scientific Reports},
month = {7},
number = {1},
pages = {6291},
pmid = {28740127},
publisher = {Springer US},
title = {{Three-Dimensional Imaging of Biological Tnumber by Cryo X-Ray Ptychography}},
url = {http://dx.doi.org/10.1038/s41598-017-05587-4 https://www.nature.com/articles/s41598-017-05587-4},
volume = {7},
year = {2017}
}

@article{Cloetens1999,
author = {Cloetens, P and Ludwig, W and Baruchel, J and {Van Dyck}, D. and {Van Landuyt}, J. and Guigay, J P and Schlenker, M},
doi = {10.1063/1.125225},
issn = {0003-6951},
journal = {Applied Physics Letters},
month = {11},
number = {19},
pages = {2912--2914},
title = {{Holotomography: Quantitative phase tomography with micrometer resolution using hard synchrotron radiation x rays}},
url = {https://pubs.aip.org/apl/article/75/19/2912/516504/Holotomography-Quantitative-phase-tomography-with},
volume = {75},
year = {1999}
}

@article{Paganin2002,
author = {Paganin, D. and Mayo, S. C. and Gureyev, T. E. and Miller, P. R. and Wilkins, S. W.},
doi = {10.1046/j.1365-2818.2002.01010.x},
issn = {0022-2720},
journal = {Journal of Microscopy},
month = {4},
number = {1},
pages = {33--40},
title = {{Simultaneous phase and amplitude extraction from a single defocused image of a homogeneous object}},
url = {https://onlinelibrary.wiley.com/doi/10.1046/j.1365-2818.2002.01010.x},
volume = {206},
year = {2002}
}

@article{Walton2015,
author = {Walton, Lucy A. and Bradley, Robert S. and Withers, Philip J. and Newton, Victoria L. and Watson, Rachel E. B. and Austin, Clare and Sherratt, Michael J.},
doi = {10.1038/srep10074},
issn = {2045-2322},
journal = {Scientific Reports},
month = {5},
number = {1},
pages = {10074},
pmid = {25975937},
publisher = {Nature Publishing Group},
title = {{Morphological Characterisation of Unstained and Intact Tnumber Micro-architecture by X-ray Computed Micro- and Nano-Tomography}},
url = {https://www.nature.com/articles/srep10074},
volume = {5},
year = {2015}
}

@inproceedings{Nathansen2024,
author = {Nathansen, Andrea and Clausen, Matthis and Berning, Manuel and MacKenzie, Ethan and Zhang, Yuxin and Pacureanu, Alexandra and Schaefer, Andreas T. and Rzepka, Norman and Bosch, Carles},
booktitle = {Developments in X-Ray Tomography XV},
doi = {10.1117/12.3028309},
editor = {M{\"{u}}ller, Bert and Wang, Ge},
isbn = {9781510679641},
issn = {1996756X},
month = {10},
pages = {46},
publisher = {SPIE},
title = {{Cell nuclei segmentation in mm-scale x-ray holographic nanotomography images of mouse brain tnumber}},
url = {https://www.spiedigitallibrary.org/conference-proceedings-of-spie/13152/3028309/Cell-nuclei-segmentation-in-mm-scale-x-ray-holographic-nanotomography/10.1117/12.3028309.full},
volume = {13152},
year = {2024}
}

@article{Varga2015,
author = {Varga, Peter and Hesse, Bernhard and Langer, Max and Schrof, Susanne and M{\"{a}}nnicke, Nils and Suhonen, Heikki and Pacureanu, Alexandra and Pahr, Dieter and Peyrin, Fran{\c{c}}oise and Raum, Kay},
doi = {10.1007/s10237-014-0601-9},
issn = {16177940},
journal = {Biomechanics and Modeling in Mechanobiology},
number = {2},
pages = {267--282},
pmid = {25011566},
title = {{Synchrotron X-ray phase nano-tomography-based analysis of the lacunar–canalicular network morphology and its relation to the strains experienced by osteocytes in situ as predicted by case-specific finite element analysis}},
volume = {14},
year = {2015}
}

@article{Nopens2025,
author = {Nopens, Martin and Greving, Imke and Flenner, Silja and Hesse, Linnea and L{\"{u}}dtke, Jan and Altgen, Michael and Koch, Gerald and Beruda, Johannes and Heldner, Sabrina and K{\"{o}}hm, Hannes and Kaschuro, Sergej and Olbrich, Andrea and Mietner, Jakob Benedikt and Scheckenbach, Fabian and Sieburg-Rockel, J{\"{o}}rdis and Krause, Andreas},
doi = {10.1107/S1600577525006484},
isbn = {1600577525006},
issn = {1600-5775},
journal = {Journal of Synchrotron Radiation},
month = {9},
number = {5},
pages = {1354--1360},
pmid = {40824694},
title = {{Design and implementation of a climate chamber for moisture sensitive nanotomography of biological samples}},
url = {https://journals.iucr.org/paper?S1600577525006484},
volume = {32},
year = {2025}
}

@article{Ulrich2025,
author = {Ulrich, Kim and Scheckenbach, Fabian and Wong, Tak Ming and Masselter, Tom and Flenner, Silja and Visconti, Anaclara and Nopens, Martin and Krause, Andreas and Kaschuro, Sergej and Mietner, Jakob Benedikt and Speck, Thomas and Greving, Imke and Zeller-Plumhoff, Berit and Hesse, Linnea},
doi = {10.3389/fpls.2025.1572745},
issn = {1664462X},
journal = {Frontiers in Plant Science},
number = {August},
pages = {1--16},
title = {{Quantifying hygroscopic deformation in lignocellulosic tnumbers: a digital volume correlation tool comparison}},
volume = {16},
year = {2025}
}

@article{Reimers2023,
author = {Reimers, Jan and Trinh, Huu Ch{\'{a}}nh and Wiese, Bj{\"{o}}rn and Meyer, Sebastian and Brehling, Jens and Flenner, Silja and Hagemann, Johannes and Kruth, Maximilian and Kibkalo, Lidia and {\'{C}}wieka, Hanna and Hindenlang, Birte and Lipinska-Chwalek, Marta and Mayer, Joachim and Willumeit-R{\"{o}}mer, Regine and Greving, Imke and Zeller-Plumhoff, Berit},
doi = {10.1021/acsami.3c04054},
issn = {19448252},
journal = {ACS Applied Materials and Interfaces},
number = {29},
pages = {35600--35610},
pmid = {37459562},
title = {{Development of a Bioreactor-Coupled Flow-Cell Setup for 3D In Situ Nanotomography of Mg Alloy Biodegradation}},
volume = {15},
year = {2023}
}

@article{Lombardo2012,
author = {Lombardo, Jeffrey J. and Ristau, Roger A. and Harris, William M. and Chiu, Wilson K. S.},
doi = {10.1107/S0909049512027252},
isbn = {0909049512027},
issn = {0909-0495},
journal = {Journal of Synchrotron Radiation},
month = {9},
number = {5},
pages = {789--796},
pmid = {22898959},
publisher = {International Union of Crystallography},
title = {{Focused ion beam preparation of samples for X-ray nanotomography}},
url = {https://journals.iucr.org/paper?S0909049512027252},
volume = {19},
year = {2012}
}

@article{Odstrcil2019,
author = {Odstr{\v{c}}il, Michal and Lebugle, Maxime and Guizar-Sicairos, Manuel and David, Christian and Holler, Mirko},
doi = {10.1364/OE.27.014981},
issn = {1094-4087},
journal = {Optics Express},
month = {5},
number = {10},
pages = {14981},
pmid = {31163938},
title = {{Towards optimized illumination for high-resolution ptychography}},
url = {https://opg.optica.org/abstract.cfm?URI=oe-27-10-14981},
volume = {27},
year = {2019}
}

@article{Mayr2021,
   author = {Sina Mayr and Simone Finizio and Joakim Reuteler and Stefan Stutz and Carsten Dubs and Markus Weigand and Aleš Hrabec and Jörg Raabe and Sebastian Wintz},
   doi = {10.3390/cryst11050546},
   issn = {2073-4352},
   number = {5},
   journal = {Crystals},
   month = {5},
   pages = {546},
   publisher = {MDPI AG},
   title = {Xenon Plasma Focused Ion Beam Milling for Obtaining Soft X-ray Transparent Samples},
   volume = {11},
   year = {2021}
}

@article{Singh2023,
   author = {Kritika Singh and Surya Snata Rout and Christina Krywka and Anton Davydok},
   doi = {10.3390/ma16227220},
   issn = {1996-1944},
   number = {22},
   journal = {Materials},
   month = {11},
   pages = {7220},
   publisher = {Multidisciplinary Digital Publishing Institute},
   title = {Local Structural Modifications in Metallic Micropillars Induced by Plasma Focused Ion Beam Processing},
   volume = {16},
   url = {https://www.mdpi.com/1996-1944/16/22/7220},
   year = {2023}
}

@article{Wolff2018,
   author = {A. Wolff and N. Klingner and W. Thompson and Y. Zhou and J. Lin and Y.Y. Peng and J.A.M. Ramshaw and Y. Xiao},
   doi = {10.1111/jmi.12731},
   issn = {0022-2720},
   number = {1},
   journal = {Journal of Microscopy},
   month = {10},
   pages = {47-59},
   pmid = {30019759},
   publisher = {John Wiley and Sons Inc},
   title = {Modelling of focused ion beam induced increases in sample temperature: a case study of heat damage in biological samples},
   volume = {272},
   year = {2018}
}

@article{John2026Adv,
   author = {Dominik John and David M. Paganin and Marie‐Christine Zdora and Lisa Marie Petzold and Patrick Ilg and Junan Chen and Sara Baggio and Johannes B. Thalhammer and Sami Wirtensohn and Julian Moosmann and Jörg U. Hammel and Felix Beckmann and Samantha J. Alloo and Jannis N. Ahlers and Madleen Busse and Julia Herzen and Kaye S. Morgan},
   doi = {10.1002/advs.202519783},
   issn = {2198-3844},
   number = {21},
   journal = {Advanced Science},
   month = {4},
   pages = {e19783},
   publisher = {John Wiley \& Sons, Ltd},
   title = {Quantitative Stain Mapping in X‐Ray Virtual Histology},
   volume = {13},
   url = {https://advanced.onlinelibrary.wiley.com/doi/10.1002/advs.202519783 http://arxiv.org/abs/2509.14768},
   year = {2026}
}

@article{Zernike1942,
   author = {F. Zernike},
   doi = {10.1016/S0031-8914(42)80035-X},
   issn = {00318914},
   number = {7},
   journal = {Physica},
   month = {7},
   pages = {686-698},
   title = {Phase contrast, a new method for the microscopic observation of transparent objects},
   volume = {9},
   url = {https://linkinghub.elsevier.com/retrieve/pii/S003189144280035X},
   year = {1942}
}

@article{Schmahl1994,
   author = {G. Schmahl and D. Rudolph and P. Guttmann and G. Schneider and J. Thieme and B. Niemann and T. Wilhein},
   doi = {10.1080/08940889408261282},
   issn = {0894-0886},
   number = {4},
   journal = {Synchrotron Radiation News},
   month = {7},
   pages = {19-22},
   publisher = {Taylor \& Francis Group},
   title = {Phase contrast X-ray microscopy},
   volume = {7},
   url = {http://www.tandfonline.com/doi/abs/10.1080/08940889408261282},
   year = {1994}
}

@article{Neuhausler2003,
   author = {U. Neuhäusler and G. Schneider and W. Ludwig and D. Hambach},
   doi = {10.1051/jp4:20030145},
   issn = {1155-4339},
   journal = {Journal de Physique IV (Proceedings)},
   month = {3},
   pages = {567-570},
   title = {Phase contrast X-ray microscopy at 4 keV photon energy with 60 nm resolution},
   volume = {104},
   url = {http://www.edpsciences.org/10.1051/jp4:20030145},
   year = {2003}
}

@article{Andrews2008,
   author = {J. C. Andrews and S. Brennan and C. Patty and K. Luening and P. Pianetta and E. Almeida and M. C. H. van der Meulen and M. Feser and J. Gelb and J. Rudati and A. Tkachuk and W. B. Yun},
   doi = {10.1080/08940880802123043},
   issn = {0894-0886},
   number = {3},
   journal = {Synchrotron Radiation News},
   month = {6},
   pages = {17-26},
   title = {A High Resolution, Hard X-ray Bio-imaging Facility at SSRL},
   volume = {21},
   url = {http://www.tandfonline.com/doi/abs/10.1080/08940880802123043},
   year = {2008}
}

@article{Stampanoni2010,
   author = {Marco Stampanoni and Rajmund Mokso and Federica Marone and Joan Vila-Comamala and Sergey Gorelick and Pavel Trtik and Konstantin Jefimovs and Christian David},
   doi = {10.1103/PhysRevB.81.140105},
   issn = {1098-0121},
   number = {14},
   journal = {Physical Review B},
   month = {4},
   pages = {140105},
   title = {Phase-contrast tomography at the nanoscale using hard x rays},
   volume = {81},
   url = {https://link.aps.org/doi/10.1103/PhysRevB.81.140105},
   year = {2010}
}

@article{Storm2020,
   author = {Malte Storm and Florian Döring and Shashidhara Marathe and Christian David and Christoph Rau},
   doi = {10.1017/S0885715620000238},
   issn = {0885-7156},
   number = {S1},
   journal = {Powder Diffraction},
   month = {12},
   pages = {S8-S14},
   publisher = {Cambridge University Press},
   title = {The Diamond I13 full-field transmission X-ray microscope: a Zernike phase-contrast setup for material sciences},
   volume = {35},
   url = {https://www.cambridge.org/core/product/identifier/S0885715620000238/type/journal_article},
   year = {2020}
}

@article{Vartiainen2014,
   author = {Ismo Vartiainen and Rajmund Mokso and Marco Stampanoni and Christian David},
   doi = {10.1364/OL.39.001601},
   issn = {0146-9592},
   number = {6},
   journal = {Optics Letters},
   month = {3},
   pages = {1601},
   pmid = {24690848},
   title = {Halo suppression in full-field x-ray Zernike phase contrast microscopy},
   volume = {39},
   url = {https://opg.optica.org/abstract.cfm?URI=ol-39-6-1601},
   year = {2014}
}

@inproceedings{Yu2017,
   author = {Boliang Yu and Loriane Weber and Alexandra Pacureanu and Max Langer and Cecile Olivier and Peter Cloetens and Francoise Peyrin},
   doi = {10.1109/ISBI.2017.7950467},
   isbn = {978-1-5090-1172-8},
   booktitle = {2017 IEEE 14th International Symposium on Biomedical Imaging (ISBI 2017)},
   month = {4},
   pages = {56-59},
   publisher = {IEEE},
   title = {Phase retrieval in 3D X-ray magnified phase nano CT: Imaging bone tnumber at the nanoscale},
   url = {http://ieeexplore.ieee.org/document/7950467/},
   year = {2017}
}

@article{Monaco2022,
   author = {F. Monaco and M. Hubert and J.C. Da Silva and V. Favre-Nicolin and D. Montinaro and P. Cloetens and J. Laurencin},
   doi = {10.1016/j.matchar.2022.111834},
   issn = {10445803},
   number = {February},
   journal = {Materials Characterization},
   month = {5},
   pages = {111834},
   publisher = {Elsevier Inc.},
   title = {A comparison between holographic and near-field ptychographic X-ray tomography for solid oxide cell materials},
   volume = {187},
   url = {https://doi.org/10.1016/j.matchar.2022.111834 https://linkinghub.elsevier.com/retrieve/pii/S1044580322001164},
   year = {2022}
}

@article{Kalbfleisch2022,
   author = {Sebastian Kalbfleisch and Yuhe Zhang and Maik Kahnt and Khachiwan Buakor and Max Langer and Till Dreier and Hanna Dierks and Philip Stjärneblad and Emanuel Larsson and Korneliya Gordeyeva and Lert Chayanun and Daniel Söderberg and Jesper Wallentin and Martin Bech and Pablo Villanueva-Perez},
   doi = {10.1107/S1600577521012200},
   issn = {1600-5775},
   number = {1},
   journal = {Journal of Synchrotron Radiation},
   month = {1},
   pages = {224-229},
   publisher = {International Union of Crystallography},
   title = {X-ray in-line holography and holotomography at the NanoMAX beamline},
   volume = {29},
   url = {https://journals.iucr.org/paper?S1600577521012200},
   year = {2022}
}

@article{Nikitin2024,
   author = {Viktor Nikitin and Marcus Carlsson and Doğa Gürsoy and Rajmund Mokso and Peter Cloetens},
   doi = {10.1364/OE.537341},
   issn = {1094-4087},
   number = {23},
   journal = {Optics Express},
   month = {11},
   pages = {41905},
   title = {X-ray nano-holotomography reconstruction with simultaneous probe retrieval},
   volume = {32},
   url = {https://opg.optica.org/abstract.cfm?URI=oe-32-23-41905},
   year = {2024}
}

@article{Lucht2025,
   author = {Jens Lucht and Paul Meyer and Leon Merten Lohse and Tim Salditt},
   doi = {10.1107/S1600577525008550},
   issn = {1600-5775},
   number = {6},
   journal = {Journal of Synchrotron Radiation},
   month = {11},
   pages = {1586-1594},
   pmid = {41159837},
   publisher = {International Union of Crystallography},
   title = {HoToPy: a toolbox for X-ray holo-tomography in Python},
   volume = {32},
   url = {https://journals.iucr.org/paper?S1600577525008550},
   year = {2025}
}

@article{Walsh2021,
   author = {C. L. Walsh and P. Tafforeau and W. L. Wagner and D. J. Jafree and A. Bellier and C. Werlein and M. P. Kühnel and E. Boller and S. Walker-Samuel and J. L. Robertus and D. A. Long and J. Jacob and S. Marussi and E. Brown and N. Holroyd and D. D. Jonigk and M. Ackermann and P. D. Lee},
   doi = {10.1038/s41592-021-01317-x},
   issn = {1548-7091},
   number = {12},
   journal = {Nature Methods},
   month = {12},
   pages = {1532-1541},
   pmid = {34737453},
   publisher = {Nature Publishing Group},
   title = {Imaging intact human organs with local resolution of cellular structures using hierarchical phase-contrast tomography},
   volume = {18},
   url = {https://www.nature.com/articles/s41592-021-01317-x},
   year = {2021}
}

@article{Twengstrm2022,
   author = {William Twengström and Carlos F. Moro and Jenny Romell and Jakob C. Larsson and Ernesto Sparrelid and Mikael Björnstedt and Hans M. Hertz},
   doi = {10.1117/1.JMI.9.3.031503},
   isbn = {9781510645189},
   issn = {2329-4302},
   number = {03},
   journal = {Journal of Medical Imaging},
   month = {2},
   pages = {031503},
   publisher = {SPIE},
   title = {Can laboratory x-ray virtual histology provide intraoperative 3D tumor resection margin assessment?},
   volume = {9},
   url = {https://www.spiedigitallibrary.org/journals/journal-of-medical-imaging/volume-9/number-03/031503/Can-laboratory-x-ray-virtual-histology-provide-intraoperative-3D-tumor/10.1117/1.JMI.9.3.031503.full},
   year = {2022}
}

@article{Schaeper2025,
   author = {Jannis J. Schaeper and Christoph A. Kampshoff and Bettina J. Wolf and Lennart Roos and Susann Michanski and Torben Ruhwedel and Marina Eckermann and Alexander Meyer and Marcus Jeschke and Carolin Wichmann and Tobias Moser and Tim Salditt},
   doi = {10.1038/s41598-025-89431-0},
   issn = {2045-2322},
   number = {1},
   journal = {Scientific Reports},
   month = {3},
   pages = {7933},
   pmid = {40050327},
   publisher = {Nature Publishing Group},
   title = {3D virtual histology of rodent and primate cochleae with multi-scale phase-contrast X-ray tomography},
   volume = {15},
   url = {https://www.nature.com/articles/s41598-025-89431-0},
   year = {2025}
}

@article{Baggio2026,
   author = {Sara Baggio and Sami Wirtensohn and Sara Savatovic and Imke Greving and Silja Flenner and Julia Herzen},
   doi = {10.1364/opticaopen.33145250},
   journal = {Optica Open},
   month = {8},
   title = {Stepping parameter optimization for near-field ptychography},
   url = {https://preprints.opticaopen.org/articles/preprint/Stepping_parameter_optimization_for_near-field_ptychography/33145250},
   year = {2026}
}

\newpage
\section{Supplements}

\subsection{Resolution}
\begin{figure}[htbp]
    \centering
    \includegraphics[width=\linewidth]{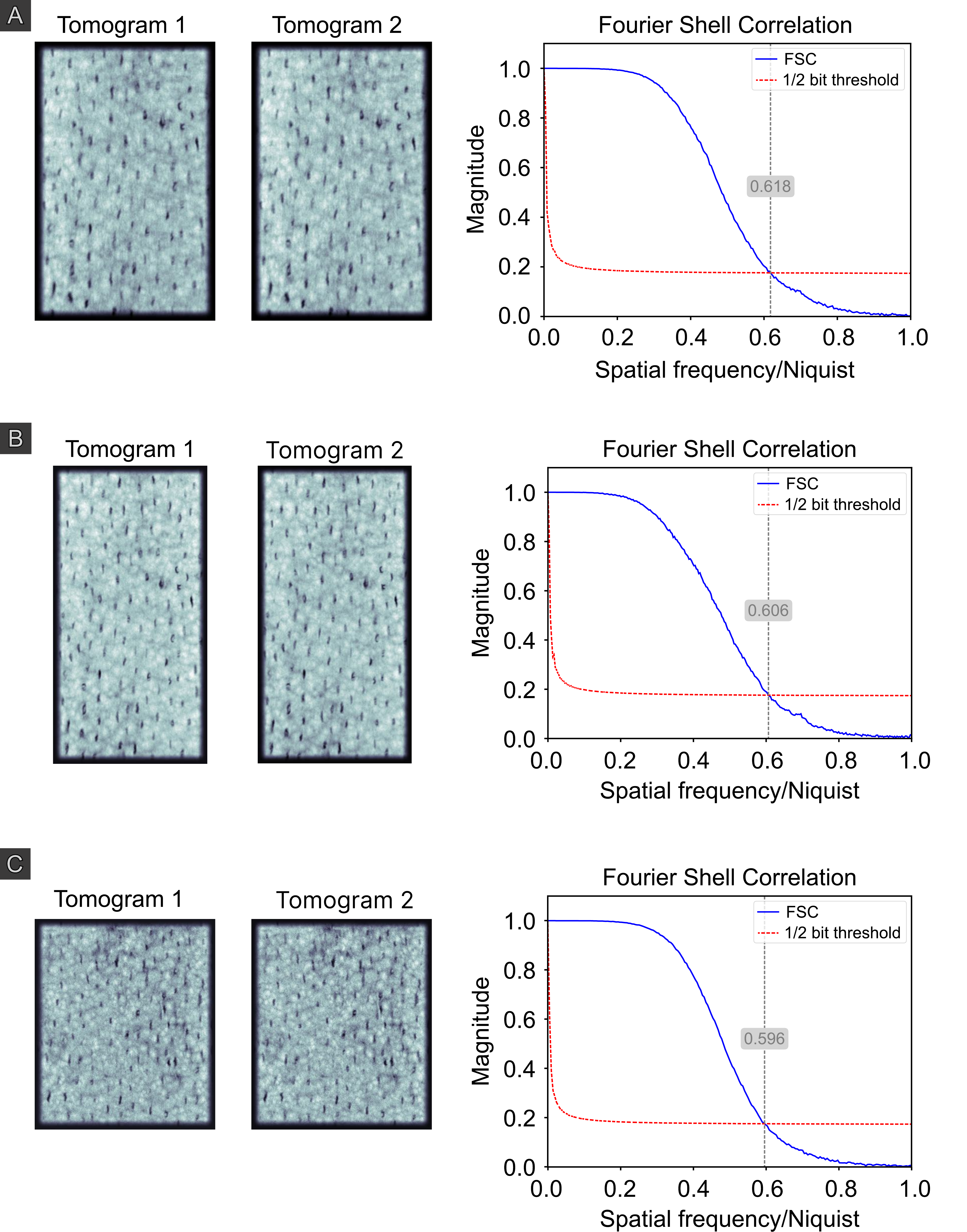}
    \caption{Results for the Fourier shell correlation of the volumes for the elephant ivory measured by the step-rotation scan (\textbf{A}), the fly-rotation scan (\textbf{B}), and the fly-rotation II scan, with reduced exposure time and less projections (\textbf{C}). The data sets are split by odd and even projections into two subsets, which are then correlated after separate reconstruction. The intersection point of the correlation function with the half-bit threshold gives the resolution by multiplying its inverse with the effective pixel size.}
    \label{fig:fsc_result}
\end{figure}

The resolution is derived by the Fourier shell correlation implemented in \textit{toupy}~\cite{daSilvia2017}. The retrieved phase projections are split into two subsets by odd and even index numbers. Each subset is reconstructed separately with the same parameters as the full tomogram, shown in the main text body. The two independent tomograms are then fed into \textit{toupy}. Figure \ref{fig:fsc_result} A, B, and C show the results for the elephant ivory measured with step-rotation, fly-rotation, and fly-rotation II with reduced exposure time and fewer projections, respectively.
\newpage

\end{document}